\documentclass[draftcls,onecolumn]{IEEEtran}

\usepackage[]{graphicx,times,epsfig,amssymb}
\usepackage[cmex10]{amsmath}
\usepackage{array}  % extended styles for tables
\usepackage{theorem}    % mathematical stuff
\usepackage{subfigure} 
\newtheorem{remark}{Remark}
\newtheorem{proposition}{Proposition}
\newtheorem{corollary}{Corollary}
\newtheorem{lemma}{Lemma}

\ifCLASSINFOpdf
\else
\fi

\def\x{{\mathbf x}}
\def\y{{\mathbf y}}
\def\z{{\mathbf z}}
\def\u{{\mathbf u}}
\def\v{{\mathbf v}}

\def\r{{\mathbf r}}

\def\m{{\mathbf m}}
\def\F{{\mathbf F}}
\def\Q{{\mathbf Q}}
\def\P{{\mathbf P}}
\def\R{{\mathbf R}}

\def\H{{\mathbf H}}
\def\B{{\mathbf B}}
\newcommand{\RR}{{\rm I\!R}}

\begin{document}
\title{A class of fast exact Bayesian filters \\ in dynamical models with jumps}

\author{Yohan~Petetin,
        Fran\c{c}ois~Desbouvries %,~\IEEEmembership{Senior Member,~IEEE}
       % <-this % stops a space
\thanks{
Yohan Petetin and 
Fran\c{c}ois Desbouvries
are with Mines Telecom Institute, Telecom SudParis, CITI Department, 
9 rue Charles Fourier, 91011 Evry, France and with CNRS UMR 5157.
We would like to thank the French MOD DGA/MRIS for financial support of the Ph.D. of Y. Petetin.}}
\maketitle

\begin{abstract}
In this paper, we focus on
the statistical filtering problem in dynamical models with jumps.
When a particular application relies on 
physical properties which are modeled by linear and Gaussian 
probability density functions with jumps, an usual
method consists in approximating the
optimal Bayesian estimate 
(in the sense of the Minimum Mean Square Error (MMSE))
in a linear and Gaussian Jump Markov State Space System (JMSS).
Practical solutions include algorithms based on numerical
approximations or based on Sequential
Monte Carlo (SMC) methods.
In this paper, we propose a class of alternative methods 
which consists in building statistical models which share
the same physical properties of interest but in which the
computation of the optimal MMSE estimate can be done at a computational
cost which is linear in the number
of observations.
%%Consequently, the algorithms obtained
%%can be used as an alternative to classical approximations. 
%%In particular, in linear and Gaussian
%%Jump Markov State Space Systems (JMSS) the computation
%%of the optimal Bayesian estimate (in the sense of the Mean Square Error (MSE))
%%is an NP hard problem. Suboptimal solutions include 
%%algorithms based on numerical approximations or based on Sequential
%%Monte Carlo (SMC) methods. We propose a class of alternative methods 
%%which consists in building statistical models which have
%%the same physical properties of a given linear and Gaussian JMSS
%%but in which the computation of the optimal estimate 
%%can be done at a computational cost which is linear in the number
%%of observations. Consequently, this class of statistical models
%%can be used to approximate the Bayesian estimate
%%in a linear and Gaussian JMSS at a linear computational cost.
\end{abstract}


% \fi
\IEEEpeerreviewmaketitle

\section{Introduction}

\subsection{Background}
\label{section-background}

\IEEEPARstart{L}{et}
%Markov State Space Systems (JMSS)
%are widely used in signal processing applications such that 
%single- or multi-target tracking, 
%finance and geology 
%\cite{Cappeetal} 
%\cite{doucet-jump-Mkv} 
%\cite{auxiliary} 
%\cite{MAHLER_LIVRE2007} 
%\cite{fernhead-jmss}
%\cite{pasha-jm-phd}.
%Their success in Bayesian inference is due to 
%their capacity to describe a probabilistic
%representation of a given physical problem
%and the possibility that they offer to 
%approximate quantities of practical interest
%when the goal is to estimate hidden data from observations.
$\{{\bf y}_k\}_{k \geq 0} \in \mathbb{R}^p$ be a  sequence of observations
and
$\{{\bf x}_k\}_{k \geq 0} \in \mathbb{R}^m$ a sequence of hidden states
(as far as notations are concerned,
%upper case letters denote random variables (r.v.),
%lower case ones their realizations,
we do not differ random variables (r.v). and
their realizations;
bold letters denote vectors;
$p(\x)$, say, denotes the probability density function (pdf)
of r.v. ${\bf x}$ and 
$p(\x|\y)$, say, the conditional pdf of ${\bf x}$ given $\y$). 
%%%%%%%%%%%%%%%%%%%%%%%%%%%%we may note couple $({\bf x},{\bf y})= {\bf z}$).
Let $\x_{0:k}=\{\x_i\}_{i=0}^k$ and
$\y_{0:k}=\{\y_i\}_{i=0}^k$.
In this paper we address the Bayesian filtering problem 
which consists in computing (an approximation of) $p(\x_n|\y_{0:n})$
and next in computing a moment of this pdf.
In this paper we directly focus on the recursive computation of 
\begin{equation}
\label{Theta_n-def-generale}
\Phi_k = {\rm E} (f(\x_k)|\y_{0:k}) =  \int f({\bf x}_{k}) p({\bf x}_{k}|{\bf y}_{0:k}) {\rm d}{\bf x}_{k} \text{,}
\end{equation}  
where $f(\x)=\x$ or $f(\x)=\x \x^T$.

Computing $\Phi_k$ is of interest in
many applications such that
single- \cite{doucet-jump-Mkv} \cite{auxiliary} \cite{MAHLER_LIVRE2007}
or multi-target tracking \cite{pasha-jm-phd}, 
finance \cite{Cappeetal} \cite{auxiliary}
and geology  \cite{fernhead-jmss}.
These applications are best modeled when in addition
to $\{\x_k \}$ and $\{\y_k \}$, 
we introduce a third
sequence $\{r_k\}_{k \geq 0}$ in which $r_k \in \{1, \cdots, K \}$ is discrete 
and hidden, and models the regime switchings.
In this case, the underlying model is essentially described
by two pdfs $f_{i|i-1}(\x_i|\x_{i-1},r_i)$ and $g_i(\y_i|\x_i,r_i)$.
Pdf $f_{i|i-1}$ describes the dynamical evolution of the hidden state over time
when regime $r_i$ is known, and $g_i$ models how the observation $\y_i$ is produced from state $\x_i$ under regime $r_i$.
From now on, we will say that pdfs $f_{i|i-1}$ and $g_i$ characterize the physical properties
of the problem at hand since they should be chosen in accordance with the considered application.
%%Physical applications %which use a Bayesian formulation
%%such that single- or multi-target tracking, 
%%finance and geology 
%%\cite{Cappeetal} 
%%\cite{doucet-jump-Mkv} 
%%\cite{auxiliary} 
%%\cite{MAHLER_LIVRE2007} 
%%\cite{fernhead-jmss}
%%\cite{pasha-jm-phd}
%%rely, for $1 \leq i \leq k$, on the transition probabilies $p(r_i|r_{i-1})$ of 
%%the jumps, on pdfs $f_{i|i-1}(\x_i|\x_{i-1},r_i)$ 
%%which describe the dynamical evolution of the hidden state over time
%%when regime $r_i$ is known and on $g_i(\y_i|\x_i,r_i)$ 
%%which model how the observation $\y_i$ is produced from state $\x_i$ under regime $r_i$.
%%From now on, we will say that pdfs $p(r_i|r_{i-1})$, $f_{i|i-1}$ and $g_i$ characterize the physical properties 
%%since they should be chosen according to a given practical problem.

One should still specify the joint probability model
for $\{\x_k,\y_k,r_k\}_{k \geq 0}$.
A popular model which is directly built from 
the pdfs of interest $f_{i|i-1}$ and $g_i$
is the so-called JMSS, 
%%%in the context of so-called
%%%JMSS
%Jump Markov State Space Systems (JMSS),
i.e. a model where the joint pdf of 
$(\x_{0:k},\y_{0:k},\r_{0:k})$ %=(\z_{0:k},\r_{0:k})$
reads
\begin{align}
\label{loi-jmss}
& p^1(\x_{0:k},\y_{0:k},\r_{0:k})
=
\underbrace{p^1(r_0)\prod_{i=1}^k p^1(r_i|r_{i-1})}_{p^1(\r_{0:k})}
%%\times \nonumber \\
%%&
\underbrace{ p^1(\x_0|r_0)\prod_{i=1}^k f_{i|i-1}(\x_i|\x_{i-1},r_i)}_{p^1(\x_{0:k}|\r_{0:k})}
\underbrace{\prod_{i=0}^n g_i(\y_i|\x_i,r_i)}_{p^1(\y_{0:k}|\x_{0:k},\r_{0:k})} .
\end{align}
This model is popular because it directly takes into account 
the physical properties of 
interest, and it reduces to the well known Hidden Markov Chain (HMC) model when the jumps 
$\r_{0:k}$ are fixed. Note that in this model, we assume that the jumps are a Markov chain
(MC).
Unfortunately, computing $\Phi_k$ in a JMSS model 
%in \eqref{Theta_n-def-generale}
is impossible in the general case,
i.e. when $f_{i|i-1}$ and $g_i$ are arbitrary functions,
and is still NP-hard in the linear and Gaussian case \cite{tugnait},
i.e. when functions $f_{i|i-1}$ and $g_i$ satisfy 
\begin{align}
\label{transition-jmss}
f_{i|i-1}(\x_i|\x_{i-1},r_i)&=\mathcal{N}(\x_i;\F_i(r_i)\x_{i-1};\Q_i(r_i)) \text{,} \\
\label{vraisemblance-jmss}
g_{i}(\y_i|\x_{i},r_i)&=\mathcal{N}(\y_i;\H_i(r_i)\x_{i};\R_i(r_i))
\end{align} 
($\mathcal{N}(\x;{\bf m};{\bf P})$ is the Gaussian pdf 
with mean ${\bf m}$ and covariance matrix ${\bf P}$
taken at point $\x$).
%$p^1(\x_k|\y_{0:k})$ is a Gaussian Mixture (GM)
%which grows exponentially with time,
%so even in this case 
From now on we focus on the linear and Gaussian case,
since even in this case approximations are necessary.
%A number of approximations have been developed.
A number of suboptimal methods for computing 
$\Phi_k$ in linear and Gaussian JMSS have been proposed so far.
First, based on the observation that
$p^1(\x_k|\y_{0:k})$ is a Gaussian Mixture (GM)
which grows exponentially with time,
numerical approximations such as pruning and merging 
have been studied \cite{tugnait} \cite{Sorenson-gaussian}.
A second class of approximations is given
by the Interacting Multiple Model (IMM) 
\cite{Shalom-IMM} 
\cite{Shalom-IMM-survey} 
\cite{rong-li-survey};
roughly speaking, a bank of Kalman Filters (KF) are used for each mode
$r_k$ and their outputs are combined according to the  
parameters of the model and to the available observations.
Finally, a more recent class of methods is based on the use of Monte Carlo 
samples and Particle Filtering (PF)
\cite{doucet-jump-Mkv} \cite{Andrieux-JM}
\cite{doucet-sequentialMC} \cite{Shon-MPF}.
A set of weighted random samples
$\{\r_{0:k}^i,w_k^i\}_{i=1}^N$ approximates
$p^1(\r_{0:k}|\y_{0:k})$,
while
$p^1(\x_{0:k}|\r_{0:k},$ $\y_{0:k})$ is a Gaussian pdf
computable via KF,
which leads to the following approximation of the pdf of $\x_{0:k}$ given $\y_{0:k}$: 
\begin{align}
p^1(\x_{0:k}|\y_{0:k}) \! & \approx \! \sum_{i=1}^N w_k(\r_{0:k}^i) \mathcal{N}(\x_{0:k};\m(\r_{0:k}^i);\P(\r_{0:k}^i)) \text{.}
\end{align}
Monte Carlo methods have suitable asymptotical convergence
properties \cite{Crisan-convergence} \cite{Cappeetal} \cite{chopin}
%(i.e. when the number of samples tend to infinity) 
but may require a serious computational cost,
since at least a KF is computed for each particle
(one has to compute $\m(\r_{0:k}^i)$ and $\P(\r_{0:k}^i)$),
and for the computation of weights
$\{w_k(\r_{0:k}^i)\}_{i=1}^N$.

\subsection{Contributions of this paper}

Let us now turn to the contents of this paper.
We assume that a particular
set of physical linear and Gaussian properties is given, i.e. 
%$p(r_0)$, $p(\x_0|r_0)$,
we are given
$p^1(r_k|r_{k-1})$, $f_{k|k-1}(\x_k|\x_{k-1},r_k)$ and $g_k(\y_k|\x_k,r_k)$.
We want to build a class of
statistical models $p^2(\x_{0:k},\y_{0:k},\r_{0:k})$
which also conveniently take into account these physical properties
(in a sense to be specified below),
but in which $\Phi_k$ can now be computed exactly and efficiently.

More precisely, our problem can be formulated as follows.
Assume
%that $p^1(r_i|r_{i-1})$, $f_{i|i-1}$ and $g_{i}$ are given,
%i.e. 
that 
\eqref{transition-jmss} and \eqref{vraisemblance-jmss}
efficiently model some practical problem of interest.
Then we look for a joint pdf $p^2(\x_{0:n},\y_{0:n},\r_{0:n})$ 
such that:
\begin{enumerate}
\item
$p^2(\x_i|\x_{i-1},r_{i})=$ $f_{i|i-1}(\x_i|$ $\x_{i-1},r_i)$;
\item
$p^2(\y_i|\x_{i},r_{i})=$ $g_{i}(\y_i|\x_{i},r_i)$; {\sl and}
\item
%%%%%%%%%%%%the exact computation of $\Phi_k$ is possible
%%%%%%%%%%%%with a computational cost linear in the number of observations.
$\Phi_k$ can be computed exactly (without resorting to any numerical or Monte Carlo approximations),
with a computational cost linear in the number of observations.
\end{enumerate}
By contrast with the methods recalled in \S \ref{section-background},
we no longer try to approximate the computation of $\Phi_k$ 
in the JMSS model $p^1(.)$, but rather
compute it exactly in a class of different statistical models
$p^2(.)$ which however share with $p^1(.)$ the same physical properties.

Let us now describe the methodology that we use to build such a pdf $p^2(.)$,
in which exact computing will be possible. 
We use a two-step procedure.
%%%%%%%%%%%%The models we will consider in this paper are 
%%%%%%%%%%%%built progressively, so let us describe our methodology.
First, we fix the jumps $\r_{0:k}$ and thus only consider process $\z_{0:k}=(\x_{0:k},\y_{0:k})$. 
When the jumps are fixed, 
JMSS models reduce to  classical HMC models,
described by pdf
\begin{align}
\label{loi-hmc}
p^1(\z_{0:k})&= \underbrace{ p^1(\x_0)\prod_{i=1}^k f_{i|i-1}(\x_i|\x_{i-1})}_{p^1(\x_{0:k})}
\underbrace{\prod_{i=0}^k g_i(\y_i|\x_i)}_{p^1(\y_{0:k}|\x_{0:k})} \text{;}
\end{align}
since model \eqref{loi-hmc} is moreover linear and Gaussian, 
%it reduces to the classical state-space system where 
$\Phi_k$ can be computed exactly via the KF.
%%%%%%%%%%%%%%%%%%%%%%%%%%%%%%%%%%%%%%%%%st sp
%%%%%%%%%%%%%%%%%%%%%%%%%%%%%%%%%%%%%%%%%kf
%%%%%%%%%%%%%%%%%%%%%%%%%%%%%%%%%%%%%%%%%So in a HMC, $\{\x_k\}_{k \geq 0}$ is a MC and
%%%%%%%%%%%%%%%%%%%%%%%%%%%%%%%%%%%%%%%%%given $\x_{0:k}$,
%%%%%%%%%%%%%%%%%%%%%%%%%%%%%%%%%%%%%%%%%observations are
%%%%%%%%%%%%%%%%%%%%%%%%%%%%%%%%%%%%%%%%%independent and
%%%%%%%%%%%%%%%%%%%%%%%%%%%%%%%%%%%%%%%%%observation ${\bf y}_i$
%%%%%%%%%%%%%%%%%%%%%%%%%%%%%%%%%%%%%%%%%depends only on state $\x_i$.
%%%%%%%%%%%%%%%%%%%%%%%%%%%%%%%%%%%%%%%%%Given a linear and Gaussian HMC model in which the computation
%%%%%%%%%%%%%%%%%%%%%%%%%%%%%%%%%%%%%%%%%of $\Phi_k$ is possible via the KF,
Adapting the objectives above,
our first goal is to compute a class of statistical models $p^2(\z_{0:k})$ 
(not necessarily HMC ones)
%in which the form of $p^2(\z_{0:k})$ 
%would be more general than factorization
%\eqref{loi-hmc};
in which the practical physical properties $f_{i|i-1}(\x_i|\x_{i-1})$
and $g_i(\y_i|\x_i)$
%%of the HMC models
would be kept, i.e. 
i) $p^2(\x_i|\x_{i-1})=f_{i|i-1}(\x_i|\x_{i-1})$, 
ii) $p^2(\y_i|\x_i)=g_i(\y_i|\x_i)$, and
iii) the computation of $\Phi_k$ in \eqref{Theta_n-def-generale}
would remain possible.
Our construction is built on Pairwise Markov Chains (PMC) models
\cite{wp-pami} \cite{derrode-wp-sp}, 
which are more general statistical models than HMC ones. 
%we will 
%derive a class of linear and Gaussian PMC models 
%in which the constraints above are verified.
Next, in the particular class of PMC models obtained,
we reintroduce the jumps and so we obtain 
a class of conditional linear and Gaussian PMC
models which satisfy the physical
properties of interest $f_{i|i-1}(\x_i|\x_{i-1},r_i)$ and 
$g_i(\y_i|\x_i,r_i)$.
%%of a given linear and Gaussian JMSS.
Among these models, we discuss
on those in which 
$\{p^2(r_{k}|\y_{0:k}),{\rm E}({\bf x}_{k}|\y_{0:k},r_{k})\}_{r_k=1}^K$
can be computed recursively;
$\Phi_k$ will eventually be computed as 
\begin{align}
\label{esperance-x}
\Phi_k 
&=
\sum_{r_{k}} p^2(r_{k}|\y_{0:k}) {\rm E}({\bf x}_{k}|\y_{0:k},r_{k}).
\end{align}
Finally, it happens that the algorithm we propose computes $\Phi_k$
%%in a a model with physical properties of interest
at a linear computational
cost in the number of observations. Thus, it can be used as 
an alternative to the classical solutions recalled in \S \ref{section-background}.
%%Indeed, simulations will 
%%show that our algorithm is more efficient than PF methods since it 
%%does not rely on MC samples.

%%%Finally, we obtain a class of models which have the physical properties
%%%of interest and in which $\Phi_k$ can be computed exactly. We also show that
%%%these models can be used to approximate the computation of
%%%$\Phi_k$ in the linear and Gaussian JMSS since both models share
%%%the same physical properties.

%%%%%%%%%%%%%%%%%%%%%%%%%%%%%%%%%%%%%%%%%%%%%%%%Among these models, we discuss
%%%%%%%%%%%%%%%%%%%%%%%%%%%%%%%%%%%%%%%%%%%%%%%%on those in which the computation of 
%%%%%%%%%%%%%%%%%%%%%%%%%%%%%%%%%%%%%%%%%%%%%%%%\begin{align}
%%%%%%%%%%%%%%%%%%%%%%%%%%%%%%%%%%%%%%%%%%%%%%%%\label{esperance-x}
%%%%%%%%%%%%%%%%%%%%%%%%%%%%%%%%%%%%%%%%%%%%%%%%{\rm E} (\x_{k+1}|\y_{0:k+1})&= \sum_{r_{k+1}} p(r_{k+1}|\y_{0:k+1}) {\rm E}({\bf x}_{k+1}|\y_{0:k+1},r_{k+1})
%%%%%%%%%%%%%%%%%%%%%%%%%%%%%%%%%%%%%%%%%%%%%%%%\end{align}
%%%%%%%%%%%%%%%%%%%%%%%%%%%%%%%%%%%%%%%%%%%%%%%%can be deduced from
%%%%%%%%%%%%%%%%%%%%%%%%%%%%%%%%%%%%%%%%%%%%%%%%$p(r_{k}|\y_{0:k})$ and
%%%%%%%%%%%%%%%%%%%%%%%%%%%%%%%%%%%%%%%%%%%%%%%%${\rm E}({\bf x}_{k}|\y_{0:k},r_{k})$.
The paper is organized as follows.
%%In section \ref{content}, we briefly
%%introduce the main result of this paper.
In section \ref{section-pmc},
we first drop the jumps and build
a particular class of linear and Gaussian PMC models
which keeps the physical properties of interest.
%%a given linear and Gaussian HMC one.
Next in section \ref{section-tmc}, we reintroduce the jumps
and we address the sequential filtering problem
in such dynamical models.
So we start by generalizing linear and 
Gaussian JMSS to a class
of conditional linear and Gaussian PMC models
which keep the physical properties of interest.
Among this new class of models,
described by two parameters,
we look for those in which 
$\Phi_k$ can be computed exactly.
%%and we also discuss on the optimal tuning of the parameters in this class of models
%%when the goal is to approximate the linear and Gaussian JMSS model.
Finally, in section \ref{section-simulation}, we 
illustrate our methodology step by step on a practical 
example and we perform simulations.
%% in some linear and Gaussian JMSS.
%%%A part of this section is devoted to the approximated
%%%computation problem of $\Phi_k$ in linear and Gaussian JMSS.
Our method is compared to classical approximating techniques such as
the Sampling Importance Resampling (SIR) algorithm \cite{doucet-jump-Mkv} and
IMM algorithms \cite{Shalom-IMM}.
We end the paper with a Conclusion.

\section{A class of physically constrained PMC models}
\label{section-pmc}

In this section we drop the dependencies in the jump process $\{r_k\}_{k \geq 0}$.
So we start from physical properties $f_{i|i-1}(\x_i|\x_{i-1})$ and 
$g_i(\y_i|\x_i)$, which in turn define the %Gaussian 
HMC model $p^1(.)$ in \eqref{loi-hmc},
in which $\Phi_k$ can be computed exactly via KF
%KF is available 
since $f_{i|i-1}$ and $g_i$ are Gaussian.
Our aim here is to embed $p^1(.)$
%$p^1(\z_{0:n}) = p^{2,\Phi_0}(\z_{0:n})$
%as a particular point of 
into a broader class of models 
%$\{p^{2,\theta}(\z_{0:n}) \}_[\theta \in \Theta}$
$\{p^{2,\theta}\}_{\theta \in \Theta}$
(i.e., $p^1 = p^{2,\Theta_0}$ for some $\Theta_0$),
which all share the physical properties of the root model $p^1$
(i.e.,  
$p^{2,\theta}(\x_i|\x_{i-1})=f_{i|i-1}(\x_i|\x_{i-1})$ and 
$p^{2,\theta}(\y_i|\x_i)=g_i(\y_i|\x_i)$ for all $\theta$),
and in which $\Phi_k$ can still be computed exactly whatever $\theta$.
Such models are described in section \ref{statistical-extension-hmc},
and are indeed particular PMC models, which we briefly recall in section \ref{recall-pmc}.
%i.e. to the classical Gaussian state-space 
%we briefly recall why PMC models 
%are a statistical extension of HMC ones and
%we focus on linear and Gaussian PMC models.
%Next, given a linear and Gaussian HMC model,
%we derive a class of PMC models which satisfy the 
%physical properties of the original HMC model.
%The interest of this class of linear and Gaussian PMC
%models will appear at the next section, when we 
%will reintroduce the jumps.
The interest of family $\{p^{2,\theta}\}_{\theta \in \Theta}$
will become clear in section \ref{section-tmc}, 
when we will reintroduce the jumps.

%%We briefly recall 
%%that the statistical structure of these models
%%does not prevent from the recursive computation of $\Phi_k$
%%in \eqref{Theta_n-def-generale}. Finally, we derive 
%%a particular class of linear and Gaussian PMC models
%%which verify the same physical constraints of a given HMC one
%%and we discuss on the proximity of both models. The interest 
%%of these models is actually twofold. First, they generalize HMC
%%models while keeping their physical properties, so they may be more
%%adapted for real scenario. Secondly, the new approximation technique
%%for the filtering problem in JMSS we will propose in section \ref{section-tmc}
%%relies on this models.

\subsection{A brief review of PMC models}
\label{recall-pmc}

In the HMC model \eqref{loi-hmc}, 
it is well known that $\{\x_k\}_{k \geq 0}$ is an MC, 
and that given $\x_{0:k}$, 
observations $\{\y_i\}$ are independent with
$p^1(\y_i|\x_{0:k}) = p^1(\y_i|\x_{i}) = g_{i}(\y_i|\x_{i})$.
On the other hand, 
a PMC model is a model in which the {\sl pair} $\{\z_k=(\x_k,\y_k)\}_{k \geq 0}$ is assumed to be an MC,
i.e. a model which satisfies 
%One can check moreover that
\begin{eqnarray}
\label{pmc1}
p^2(\x_i,\y_i|\x_{0:i-1},\y_{0:i-1}) 
  & = & 
p^2_{i|i-1}(\x_i,\y_i|\x_{i-1},\y_{i-1})
\\
\label{pmc2}
 & = & 
p^2(\x_i|\z_{i-1}) p^2(\y_i|\x_{i-1:i},\y_{i-1})
\end{eqnarray}
Therefore, in a PMC model, pdf of $(\x_{0:k},\y_{0:k})$ reads
\begin{equation}
\label{loi-pmc}
p^2(\x_{0:k},\y_{0:k}) = p^2(\x_0,\y_0) \prod_{i=1}^k p_{i|i-1}(\x_i,\y_i|\x_{i-1},\y_{i-1}) \text{.}
\end{equation}
One can check easily that the HMC model is indeed one particular PMC,
because from \eqref{loi-hmc},
$p^1(\x_i,\y_i|\x_{0:i-1},$ $\y_{0:i-1}) =$
$f_{i|i-1}(\x_i|\x_{i-1})$
$g_{i}(\y_i|\x_{i})$.
So \eqref{pmc1} is satisfied,
and moreover the two factors in \eqref{pmc2} respectively reduce to 
%%%%%%%%%%%%%%%%%%%%%%%%%%%%%%%%%%%%%%%%%%%%%%%%%%%%%%%%%%%%%%%%%%%%%$p^1(\x_i|\x_{i-1},$ $\y_{i-1}) = f_{i|i-1}(\x_i|\x_{i-1})$ and
%%%%%%%%%%%%%%%%%%%%%%%%%%%%%%%%%%%%%%%%%%%%%%%%%%%%%%%%%%%%%%%%%%%%%$p^1(\y_i|\x_i,\x_{i-1},\y_{i-1}) = g_{i}(\y_i|\x_{i})$.
\begin{eqnarray}
\label{cond-independence1}
p^1(\x_i|\x_{i-1},\y_{i-1})    & = & f_{i|i-1}(\x_i|\x_{i-1}), \\
\label{cond-independence2}
p^1(\y_i|\x_i,\x_{i-1},\y_{i-1})  & = & g_{i}(\y_i|\x_{i}).
\end{eqnarray}
Now in a general PMC model \eqref{pmc1} is satisfied,
but 
$p^2(\x_i|\x_{i-1},\y_{i-1})$ 
may depend on both $\x_{i-1}$ and $\y_{i-1}$, and
$p^2(\y_i|\x_i,\x_{i-1},\y_{i-1})$ 
may depend on $\x_{i}$, $\x_{i-1}$ and $\y_{i-1}$.
One can show that in a PMC model, 
$\{\x_k\}_{k \geq 0}$ is no longer necessary an MC, 
and given $\x_{0:k}$, observations $\y_i$ can be dependent \cite{nsip2003}.

%%So PMC models are characterized by 
%%their initial pdf $p^2_0(\x_0,\y_0)$
%%and their transitions $p^2_{i|i-1}(.|.)$.
%%Moreover, according to properties 
%%\textbf{P.4}, \textbf{P.5} and \textbf{P.6},
%%a HMC is a particular PMC with 
%%\begin{align}
%%\label{hmc-pmc}
%%p^1_{k|k-1}(\x_k,\y_k|\x_{k-1},\y_{k-1})&=f_{k|k-1}(\x_k|\x_{k-1})g_{k}(\y_k|\x_{k}) \text{.}
%%\end{align}

As an illustration let us consider the classical state-space system
%A well-known example of linear and Gaussian HMC model is dwhich can be also defined from the following classical 
\begin{align}
%\x_0 & \sim \mathcal{N} (.;\m_{0};\P_{0}) \text{,} \\
\label{Gaussian-model-1}
%\x_k &=\F_k \x_{k-1} + {\bf b}_k + \u_k \text{, } \\
\x_k &=\F_k \x_{k-1} + \u_k \text{, } \\
\label{Gaussian-model-2}
%\y_k &= \H_k \x_k + {\bf c}_k + \v_k \text{, }
\y_k &= \H_k \x_k + \v_k \text{, }
\end{align}
in which 
%$\{{\bf b}_k\}_{k \geq 1}$ and
%$\{{\bf c}_k\}_{k \geq 0}$ are given constants,
$\{\u_k \sim \mathcal{N}(.;0;\Q_k) \}_{k \geq 1}$ and
$\{\v_k\sim \mathcal{N}(.;0;\R_k) \}_{k \geq 0}$ 
(in this paper, we assume that all covariance matrices
are positive definite)
are independent
%zero-mean Gaussian noise,
and independent of r.v.
$\x_0 \sim \mathcal{N}(.;\m_0;\P_0)$.
%with ${ \rm E}(\u_k\u_k^T)$ $=\Q_k$
%and ${ \rm E}(\v_k\v_k^T)$ $=\R_k$
We assume that $\F_k$ is invertible for all $k$.
%and that ${\bf b}_k={\bf 0}$
%and ${\bf c}_k={\bf 0}$.Now in this paper we only deal with Gaussian models.
Model 
\eqref{Gaussian-model-1}-\eqref{Gaussian-model-2}
is a Gaussian HMC model with
\begin{align}
\label{transition-hmc}
p^1(\x_k|\x_{k-1})  &=  f_{k|k-1}(\x_k|\x_{k-1}) =  \mathcal{N}(\x_k;\F_k \x_{k-1} ; \Q_k) \text{,} \\
\label{vraisemblance-hmc}
p^1(\y_k|\x_{k})   &=  g_{k}(\y_k|\x_{k}) =  \mathcal{N}(\y_k;\H_k \x_k  ; \R_k) \text{,}
\end{align}
%%%%%%%%%%%%%%%%%%%%%%%%%%%%%%%%%%%%%%%%%%%%%%%%%%%%%%%%%%%%%%%%%%%%%%$f_{n|n-1}(\x_n|\x_{n-1})&=\mathcal{N}(\x_n;\F_n \x_n ; \Q_n)$,
%%%%%%%%%%%%%%%%%%%%%%%%%%%%%%%%%%%%%%%%%%%%%%%%%%%%%%%%%%%%%%%%%%%%%%$g_{n|n}(\y_n|\x_{n})&=\mathcal{N}(\y_n;\H_n \x_n ; \R_n)$,
and as such is a particular PMC model,
in which the initial and transition pdfs of MC 
$\{(\x_k,\y_k)\}_{k \geq 0}$ read
\begin{align}
\label{initial-hmc-1}
p^1(\z_0)&= \mathcal{N}\left(\z_0; \begin{bmatrix} \m_{0} \\ \H_0 \m_{0} \end{bmatrix} ;
\begin{bmatrix}
\P_{0} & (\H_0 \P_{0})^T \\
\H_0 \P_{0} & \R_0 +  \H_0 \P_{0} \H_0^T 
\end{bmatrix} \right) \text{,} \\
\label{transition-hmc-2}
p^{1}_{k|k-1}(\z_k|\z_{k-1})&= 
%\nonumber \\ &
\mathcal{N}  \left(\z_k;  \begin{bmatrix} \F_k & {\bf 0} \\ \H_k \F_k & {\bf 0} \end{bmatrix}  \z_{k-1};  
\begin{bmatrix} \Q_k & (\H_k\Q_k)^T \\ \H_k \Q_k & \R_k + \H_k\Q_k\H_k^T \end{bmatrix}  \right) \text{.} 
\end{align}
This linear and Gaussian HMC model
\eqref{Gaussian-model-1}-\eqref{Gaussian-model-2}
(or equivalently \eqref{initial-hmc-1}-\eqref{transition-hmc-2})
appears as a particular model of the class of linear 
and Gaussian PMC models defined by:
\begin{align}
\label{initial-pmc}
p^2(\z_0)&=\mathcal{N}(\z_0; \m'_{0}; \P'_0) \text{,} \\
\label{transition-pmc}
p^2_{k|k-1}(\z_k|\z_{k-1})  &=  \mathcal{N}\left(\z_k; \underbrace{\begin{bmatrix} \F_k^1 & \F_k^2 \\ \H_k^1 & \H_k^2 \end{bmatrix}}_{\B_k} \z_{k-1} ;\underbrace { \begin{bmatrix} {\bf \Sigma}_k^{11} & {{\bf \Sigma}_k^{21}}^T \\ {\bf \Sigma}_k^{21} & {\bf \Sigma}_k^{22} \end{bmatrix}}_{{\bf \Sigma}_k} \right)  \text{.} 
%%%\label{modele-couple-gauss}
%%%\B_k &= \begin{bmatrix} \F_k^1 & \F_k^2 \\ \H_k^1 & \H_k^2 \end{bmatrix} \text{,} \\
%%%\label{sigma-couple}
%%%{\bf \Sigma}_k &= \begin{bmatrix} {\bf \Sigma}_k^{11} & {{\bf \Sigma}_k^{21}}^T \\ {\bf \Sigma}_k^{21} & {\bf \Sigma}_k^{22} \end{bmatrix} \text{.}
\end{align}
Finally, let us recall that 
in linear and Gaussian HMC models \eqref{initial-hmc-1}-\eqref{transition-hmc-2},
$\Phi_k$ in \eqref{Theta_n-def-generale}
can be computed via the KF,
%Although linear and Gaussian PMC models
%are more general,
and that KF is still available in linear and Gaussian PMC ones 
\cite [eqs. (13.56) and (13.57)] {lipster-shiryaev}
\cite{KalmanPairwise}.
\subsection{A class of physically constrained PMC models}
\label{statistical-extension-hmc}
%\subsubsection{A Particular class of linear and Gaussian HMC models}
%%Although the practical interest of linear and Gaussian PMC models
%%can be motivated by particular situations such that
%%of paragraph \ref{exemple-pmc},
%%we use a different approach to enlighten 
%%the importance of these models.

Remember that HMC models
enable to model many practical problems 
via functions 
$f_{k|k-1}(\x_k|\x_{k-1})$ and 
$g_{k}(\y_k|\x_{k})$,
but due to their statistical structure
\eqref{loi-hmc} 
%%prevents from the computation of 
$\Phi_k$ cannot be computed exactly when we reintroduce the jumps.
Thus, our objective is to derive a class of
models in which the physical properties are
the same but in which the statistical structure
may lead to the exact computation of $\Phi_k$.
To that end, we first derive a class of linear and 
Gaussian PMC models in which the physical properties are equivalent to
those of a given linear and Gaussian HMC model
\eqref{transition-hmc}-\eqref{vraisemblance-hmc},
but in which the statistical structure is more general.
%We will discuss in section \eqref{section-tmc} 

In a general PMC \eqref{transition-pmc},
the transition state and the likelihood pdfs read
\begin{align}
%%%p^2(\x_k|\x_{k-1})&= \frac{\int p^2(\x_0,\y_0) \prod_{i=1}^k p_{i|i-1}(\x_i,\y_i|\x_{i-1},\y_{i-1}) {\rm d} \x_{0:k-1} \y_{0:k}   }
%%%{\int p(\x_0,\y_0) \prod_{i=1}^{k-1} p_{i|i-1}(\x_i,\y_i|\x_{i-1},\y_{i-1}) {\rm d} \x_{0:k-1} \y_{0:k-1}  } \text{,} \\ 
%%%p^2(\y_k|\x_{k})&= \frac{\int p^2(\x_0,\y_0) \prod_{i=1}^k p_{i|i-1}(\x_i,\y_i|\x_{i-1},\y_{i-1}) {\rm d} \x_{0:k} \y_{0:k-1}   }
%%%{\int p(\x_0,\y_0) \prod_{i=1}^k p_{i|i-1}(\x_i,\y_i|\x_{i-1},\y_{i-1}) {\rm d} \x_{0:k-1} \y_{0:k}} \text{.}
p^2(\x_k|\x_{k-1})&= \frac{\int p^2(\z_0) \prod_{i=1}^k p_{i|i-1}^2(\z_i|\z_{i-1}) {\rm d} \x_{0:k-2} \y_{0:k}   }
{\int p^2(\z_0) \prod_{i=1}^{k-1} p_{i|i-1}^2(\z_i|\z_{i-1}) {\rm d} \x_{0:k-2,k}{\rm d} \y_{0:k}} \text{,} \\ 
p^2(\y_k|\x_{k})&= \frac{\int p^2(\z_0) \prod_{i=1}^k p_{i|i-1}^2(\z_i|\z_{i-1}) {\rm d} \x_{0:k-1} \y_{0:k-1}   }
{\int p^2(\z_0) \prod_{i=1}^k p_{i|i-1}^2(\z_i|\z_{i-1}) {\rm d} \x_{0:k-1} \y_{0:k}} \text{,}
\end{align}
but remember that $p^2(\x_k|\x_{k-1})$ can be different from 
$p^2(\x_k|\x_{0:k-1})$ and $p^2(\y_k|\x_{k})$ does not necessary
correspond to $p^2(\y_k|\y_{0:k-1},\x_{0:k})$.
Given \eqref{transition-hmc}-\eqref{vraisemblance-hmc}, 
we now build a class of PMC models $p^{2,\theta}$ such that for all $\theta$,
$p^{2,\theta}(\x_0)$, $p^{2,\theta}(\x_k|\x_{k-1})$ and 
$p^{2,\theta}(\y_k|\x_{k})$ coincide
with $p^1(\x_0)$, $f_{k|k-1}(\x_k|\x_{k-1})$
and $g_{k}(\y_k|\x_{k})$, respectively, and 
such that transition
$p^{2,\theta}_{k|k-1}(\z_k|\z_{k-1})$ does not necessary
depend on the parameters of initial pdf $p^2(\z_0)$. 
We have the following result
%%(a proof can be found in Appendix \ref{proof-proposition1}).
(a proof can be found in \cite[Appendix B] {PHD-couple}).
\begin{proposition}
\label{prop-1}
Let us consider a set of pdfs 
$p^1(\x_0)=\mathcal{N}(\x_0;\m_{0};\P_{0})$, and
$f_{k|k-1}$ and $g_k$ given by
\eqref{transition-hmc}-\eqref{vraisemblance-hmc},
for all $k$.
%%%\begin{align}
%%%\label{constraint-1}
%%%p^1(\x_0)&=\mathcal{N}(\x_0;\m_{0};\P_{0}) \text{,} \\
%%%\label{constraint-2}
%%%f_{k|k-1}(\x_k|\x_{k-1})&=\mathcal{N}(\x_k;\F_k\x_{k-1};\Q_k) \text{ and } \\
%%%\label{constraint-3}
%%%g_{k}(\y_k|\x_k)&=\mathcal{N}(\y_k;\H_k\x_{k};\R_k) \text{,}
%%%\end{align}
%%%for all $k$.
Then the class of linear and Gaussian PMC models $p^{2,\theta}(\x_{0:k},\y_{0:k})$ which satisfy the constraints
\begin{eqnarray}
\label{condition-0}
p^{2,\theta}(\x_0)&=&p^1(\x_0) \text{,} \\
\label{condition-1} 
p^{2,\theta}(\x_k|\x_{k-1}) &=& f_{k|k-1}(\x_k|\x_{k-1}) \text{,} 
\\
\label{condition-2} 
p^{2,\theta}(\y_k|\x_{k}) &=& g_{k}(\y_k|\x_k) \text{,}
\end{eqnarray}
and such that
$p^{2,\theta}_{k|k-1}(\x_k,\y_k|\x_{k-1},\y_{k-1})$ does
not depend on parameter $\m_0$,
%$(\m_0, \P_0)$
is described by the following 
equations:
\begin{align}
\label{init-prop-1}
p^{2,\theta}(\z_0)  &=  \mathcal{N}\left(\z_0 ; \begin{bmatrix} \m_{0} \\ \H_0 \m_{0} \end{bmatrix} ;
\begin{bmatrix}
\P_{0} & (\H_0 \P_{0})^T \\
\H_0 \P_{0} & \R_0 +  \H_0 \P_{0} \H_0^T 
\end{bmatrix} \right) \text{,} \\
%%\end{align}
%%\begin{align}
\label{transition-prop-1}
p^{2,\theta}_{k|k-1}(\z_k|\z_{k-1})  &=  \mathcal{N}(\z_k; \B_k \z_{k-1};
{\bf \Sigma}_k) \text{,}
\end{align}
where matrices $\B_k$ and ${\bf \Sigma}_k$
are defined by
\begin{align}
\label{matrice-B}
\B_k&=\begin{bmatrix}
\F_{k}-\F_{k}^2\H_{k-1} & \F_{k}^2 \\
\H_{k}\F_{k}-\H^2_{k}\H_{k-1} & \H^2_{k}
\end{bmatrix} \text{,} \\
%%%{\bf \Sigma_k}&=
%%%\begin{bmatrix}
%%%\Q_{k} \!- \! \F_{k}^2 \R_{k-1} {\F_{k}^2}^T & [{\H}_{k}{\Q}_{k} \!- \!\H^2_{k}{\R}_{k-1}\F^2_{k}]^T \\
%%%{\H}_{k}{\Q}_{k} \!- \! \H^2_{k}{\R}_{k-1}{\F^2_{k}}^T &  \R_{k}\!-\! \H^2_{k}\R_{k-1}{\H^2_k}^T \! + \! \H_{k}\Q_{k}\H_{k} 
%%%\end{bmatrix} \text{,} \\
\label{matrice-sigma}
{\bf \Sigma}_k&= \begin{bmatrix} {\bf \Sigma}_k^{11} & ({\bf \Sigma}_k^{21})^{T} \\ {\bf \Sigma}_k^{21} & {\bf \Sigma}_k^{22} \end{bmatrix} \text{,} \\
\label{matrice-sigma-11}
{\bf \Sigma}_k^{11} &= \Q_{k} -  \F_{k}^2 \R_{k-1} ({\F_{k}^2})^T \text{,} \\
\label{matrice-sigma-21}
{\bf \Sigma}_k^{21} &= {\H}_{k}{\Q}_{k} -  \H^2_{k}{\R}_{k-1} ({\F^2_{k}})^T \text{,} \\
\label{matrice-sigma-22}
{\bf \Sigma}_k^{22} &=\R_{k}\!-\! \H^2_{k}\R_{k-1}({\H^2_k})^T \! + \! \H_{k}\Q_{k}(\H_{k})^T \text{,} 
\end{align}
and where parameters 
$\theta = \{(\F_{k}^2,\H^2_{k})\}_{k \geq 1}$
can be arbitrarily chosen,
provided
%$\Q_{k}- \F_{k}^2 \R_k {\F_{k}^2}^T$, 
%$\R_{k}-\H^2_{k}\R_k{\H^2}^T+\H_{k}\Q_{k}\H_{k}$ and 
${\bf \Sigma}_k$ is a positive definite covariance matrix for all $k$.
\end{proposition} 

%\subsection{Discussion}
\subsection{Discussion and invariance properties}
\label{discussion}

%Although the statistical structure of PMC 
%models \eqref{init-prop-1}-\eqref{matrice-sigma-22} is
%more general than HMC ones, the class of
%linear and Gaussian PMC models derived above
%and the corresponding linear and Gaussian HMC one
%\eqref{initial-hmc-2}-\eqref{transition-hmc-2}
%share some common statistical properties
%which are illustrated with the following remarks.

Let us now discuss the properties 
of the constrained PMC models 
$\{p^{2,\theta}\}_{\theta \in \Theta}$
%(where each $\theta$ is some set $\{(\F_{k}^2,\H^2_{k})\}_{k \geq 1}$)
described by Proposition \ref{prop-1}.

\begin{proposition}
\label{remark-1}
%{\rm
Let us set $\H_k^2= \H_k \F_k^2$. 
From Lemma \ref{lemme-2}
(see Appendix \ref{appendice-lemma}),
$p^2(\y_k|\x_{k-1},\x_k,\y_k)$ reduces to $g_k(\y_k|\x_k)$.
If in addition $\F_k^2= {\bf 0}_{m \times p}$,
$p^2(\x_k|\x_{k-1},\y_{k-1})$ reduces to 
$f_{k|k-1}(\x_k|\x_{k-1})$ and in this case 
the PMC model %of Proposition \ref{prop-1}
reduces to the classical HMC model
\eqref{initial-hmc-1}-\eqref{transition-hmc-2}
(i.e.,
$p^1 = p^{2,\theta_0}$
with
$\theta_0 = \{(\F_{k}^2 = {\bf 0}_{m \times p},\H^2_{k}= {\bf 0}_{p \times p})\}_{k \geq 1}$).
%%%%%%%%%%%%%%%%%%%%%%%%%%%%%%%%%%%%%%%%%%%%%%%%%%%%%%%It is not surprising since the above
%%%%%%%%%%%%%%%%%%%%%%%%%%%%%%%%%%%%%%%%%%%%%%%%%%%%%%%class of PMC models
%%%%%%%%%%%%%%%%%%%%%%%%%%%%%%%%%%%%%%%%%%%%%%%%%%%%%%%extend that of linear and Gaussian HMC ones
%%%%%%%%%%%%%%%%%%%%%%%%%%%%%%%%%%%%%%%%%%%%%%%%%%%%%%%\eqref{initial-hmc}-\eqref{vraisemblance-hmc}.
%}
\end{proposition}
We now turn to invariance properties of family 
$\{p^{2,\theta}\}_{\theta \in \Theta}$
(proofs can be found in Appendix \ref{proof-remark-2}).
\begin{proposition}
\label{remark-2}
%{\rm 
Pdf $p^{2,\theta}(\x_k,\y_k|\x_{k-1})$ does not depend on $\theta$:
for all $\theta$,
\begin{align}
\label{loi-sachant-x-k-1}
 p^{2,\theta}(\x_k,\y_k|\x_{k-1})&=p^1(\x_k,\y_k|\x_{k-1}) 
\\ &  
=f_{k|k-1}(\x_k|\x_{k-1}) g_k(\y_k|\x_k) \text{.}
\end{align}
However, note that in an HMC $p^1(\x_k,\y_k|\x_{k-1})=$ $p^1(\x_k,$ $\y_k|\x_{k-1},\y_{k-1})$,
while in general 
$p^{2,\theta}(\x_k,$ $\y_k|\x_{k-1},$ $\y_{k-1})$ is different from $p^{2,\theta}(\x_k,$ $\y_k|\x_{k-1})$.
%}
\end{proposition}
From Proposition \ref{remark-2} 
we already know that 
$p^{2,\theta}(\y_k|\x_k,\x_{k-1})=p^{2,\theta}(\y_k|\x_{k})=g_k(\y_k|\x_k)$.
But indeed 
\begin{proposition}
\label{remark-3}
%{\rm
$p^{2,\theta}(\y_k|\x_{0:k})$ and 
$p^{2,\theta}(\x_{0:k})$ do not depend on $\theta$: for all $\theta$,
\begin{eqnarray}
%{\rm For\;all\;}\theta, 
\label{loidey_ksachantx_0:k}
p^{2,\theta}(\y_k|\x_{0:k})
& = &
p^{1}(\y_k|\x_{k})=
g_k(\y_k|\x_k),
\\
\label{p2xestmarkov}
p^{2,\theta}(\x_{0:k})
& = &
%p^{1}(\x_{0:k}) =
p^1(\x_0)\prod_{i=1}^k f_{i|i-1}(\x_i|\x_{i-1}).
\end{eqnarray}
%}
\end{proposition}
%%%%%%%%%%%%%%%%%%%%%%%%%%%%%%%%%%%%%%%%, see Appendix \ref{proof-remark-3}.
%%%%%%%%%%%%%%%%%%%%%%%%%%%%%%%%%%%%%%%So the general expression of the pdf
%%%%%%%%%%%%%%%%%%%%%%%%%%%%%%%%%%%%%%%of couples $({\bf x}_{0:k},{\bf y}_{0:k})$ which belong
%%%%%%%%%%%%%%%%%%%%%%%%%%%%%%%%%%%%%%%to the class of models derived in Proposition \ref{prop-1}
%%%%%%%%%%%%%%%%%%%%%%%%%%%%%%%%%%%%%%%reads
%%%%%%%%%%%%%%%%%%%%%%%%%%%%%%%%%%%%%%%\begin{align}
%%%%%%%%%%%%%%%%%%%%%%%%%%%%%%%%%%%%%%%p^2(\x_{0:k},\y_{0:k})&=p^1_0(\x_0)\prod_{i=1}^k f_{i|i-1}(\x_i|\x_{i-1})
%%%%%%%%%%%%%%%%%%%%%%%%%%%%%%%%%%%%%%%p^2(\y_0|\x_{0:k})\prod_{i=1}^k p^2(\y_i|\y_{i-1},\x_{i-1:k}) \text{;}
%%%%%%%%%%%%%%%%%%%%%%%%%%%%%%%%%%%%%%%\end{align}
%%%%%%%%%%%%%%%%%%%%%%%%%%%%%%%%%%%%%%%%The proof is given in Appendix \ref{proof-remark-3}.
Let us finally come to the global structure of $p^{2,\theta}(\x_{0:k},\y_{0:k})$.
From \eqref{p2xestmarkov},
whatever the model (i.e. whatever parameter $\theta$),
%$p^2(\x_{0:k})$ is invariant and thus the marginal process 
$\{ \x_k \}_{k\geq 0}$ is an MC with given pdf $p^1$.
So 
$p^{2,\theta}(\x_{0:k},\y_{0:k})$ only differs through 
$p^{2,\theta}(\y_{0:k}|\y_{0:k})$,
which in a PMC model reads:
\begin{equation}
\label{pysachantx-caspmc}
p^{2,\theta}(\y_{0:k}|\x_{0:k})
 = 
p^{2,\theta}(\y_0|\x_{0:k})  \prod_{i=1}^n p^{2,\theta}(\y_i|\y_{i-1},\x_{i-1:k}) \text{.}
\end{equation}
However, some simplifications occur in particular cases.
From Proposition \ref{remark-1},
if $\H_k^2 =\H_k\F_k^2$,
$p^{2,\theta}(\y_i|\y_{i-1},\x_{i-1:k})$
reduces to $p^{2,\theta}(\y_i|\x_{i:i+1})$.
On the other hand, 
if $\F_k^2= {\bf 0}$,
$p^{2,\theta}(\y_i|\y_{i-1},\x_{i-1:k})$ reduces to 
$p^{2,\theta}(\y_i|\y_{i-1},\x_{i-1:i})$.
Of course, if we set both 
$\H_k^2 =\H_k\F_k^2$ and $\F_k^2= {\bf 0}$,
then 
$p^{2,\theta}(\y_i|\y_{i-1},\x_{i-1:k})$ reduces to 
$p^{2,\Phi_0}(\y_i|\x_{i}) =$ 
$g_i(\y_i|\x_{i})$,
and \eqref{pysachantx-caspmc} to the conditional pdf $p^1(\y_{0:k}|\x_{0:k})$ 
of the HMC model \eqref{loi-hmc}.
\section{An exact filtering  algorithm in constrained conditional PMC models}
\label{section-tmc}
In the previous section, the jumps $\r_{0:k}$
were omitted. 
So let us now reintroduce  $\r_{0:k}$ 
%%in the HMC model \eqref{loi-hmc}, which leads
%%to the JMSS model \eqref{loi-jmss}, and 
in the PMC model \eqref{loi-pmc},
which leads to conditional PMC models 
that we will describe in this section. 
Thus, we will show that the physical properties 
described by $f_{i|i-1}(\x_i|\x_{i-1},r_i)$
and $g_i(\y_i|\x_i,r_i)$ can be kept 
in a particular class of conditional PMC models. 
Among the models which belong to this class, 
we will extract those for which the computation
of $\Phi_k$ is possible, and we will discuss 
on the consequences.

%%%Let us now reintroduce the jumps $\r_{0:k}$ 
%%%which were fixed in the previous section and thus 
%%%omitted. So in this section, we 
%%%first consider Conditional PMC models, which generalize
%%%the PMC models of section \ref{recall-pmc} in the same 
%%%way that JMSS generalize HMC models. 
%%%Next, we derive a class of conditional linear and Gaussian
%%%PMC models which keep the physical properties of a
%%%given linear and Gaussian JMSS and among these models
%%%we look for those the exact computation of \eqref{esperance-x} 
%%%is possible. Finally, we discuss on our approximation 
%%%when we approximate the original JMSS by a model
%%%in which the exact filtering is possible.

\subsection{Conditional PMC models}
Let us now consider the PMC model 
\eqref{loi-pmc} in which we add a dependency 
in a discrete MC $\{r_k\}_{k \geq 0}$,
which affects 
process $\z_{0:k}$ \cite{wp-cras-chaines3}
\cite{wp-cras-semi-Markov}
\cite{ieeetrsp-TMC}.
Pdf $p^2(\z_{0:k},\r_{0:k})$  of $(\z_{0:k},\r_{0:k})=$ 
$({\bf x}_{0:k},{\bf y}_{0:k},{\bf r}_{0:k})$ 
reads 
%%(notation $p^2(.)$ is now used for conditional PMC
%%models while $p^1(.)$ is reserved for classical JMSS)
\begin{align}
\label{loi-pmc-conditional}
& p^2({\bf z}_{0:k},{\bf r}_{0:k})=p^2(r_0) 
%\times
%%\nonumber \\ &
 \prod_{i=1}^k p^2(r_i|r_{i-1})p^2(\z_0|r_0)
\prod_{i=1}^k p^2_{i|i-1}(\z_i|\z_{i-1},\r_{i-1:i}) \text{.}
\end{align} 
Note that the above conditional PMC models
extend the classical JMSS \eqref{loi-jmss}
%%%if 
%%%$p^2(r_0)=$ $p^1(r_0)$, $p^2(r_k|r_{k-1})=$ $p^1(r_k|r_{k-1})$,
%%%$p^2(\z_0|r_0)=$ $p^1(\x_0|r_0)g_0(\y_0|\x_0,r_0)$
%%%and $p^2_{i|i-1}(\z_i|\z_{i-1},\r_{i-1:i})= $ 
%%%$f_{i|i-1}(\x_i|\x_{i-1},r_i) g_i(\y_i|\x_i,r_i)$.
and that contrary to a classical JMSS \eqref{loi-jmss}, 
given  $\z_{i-1}$, $r_{i-1}$  and $r_i$,
$\z_i$ can also depend on $r_{i-1}$.
Because we consider linear and Gaussian models in this paper,
we assume that the general form of $p^2_{k|k-1}(\z_k|\z_{k-1},$ $\r_{k-1:k})$
reads (notation $\F_k^1(.)$, say, is shorthand for $\F_k^1(\r_{k-1:k})$)
%%%\begin{align}
%%%\label{condi-pmc-gaussien}
%%%& p^2_{k|k-1}(\z_k|\z_{k-1},\r_{k-1:k})=
%%%\mathcal{N}\left(\z_k;
%%%\begin{bmatrix}
%%%\F_k^1(\r_{k-1:k}) & \F_k^2(\r_{k-1:k}) \\
%%%\H_k^1(\r_{k-1:k}) & \H_k^2(\r_{k-1:k})
%%%\end{bmatrix}\z_{k-1};
%%%\begin{bmatrix}
%%%{\bf \Sigma}_k^{11}(\r_{k-1:k}) & {{\bf \Sigma}_k^{21}(\r_{k-1:k})}^{T} \\ {\bf \Sigma}_k^{21}(\r_{k-1:k}) & {\bf \Sigma}_k^{22}(\r_{k-1:k})
%%%\end{bmatrix} \right)
%%%\text{;}
%%%\end{align}
\begin{align}
\label{condi-pmc-gaussien}
& p^2_{k|k-1}(\z_k|\z_{k-1},\r_{k-1:k})=
%\nonumber \\ &
\mathcal{N}\left(\z_k;
\begin{bmatrix}
\F_k^1(.) & \F_k^2(.) \\
\H_k^1(.) & \H_k^2(.)
\end{bmatrix}\z_{k-1};
\begin{bmatrix}
{\bf \Sigma}_k^{11}(.) & {{\bf \Sigma}_k^{21}(.)}^{T} \\ {\bf \Sigma}_k^{21}(.) & {\bf \Sigma}_k^{22}(.)
\end{bmatrix} \right)
\text{;}
\end{align}
note that the particular setting
$\F_k^1(\r_{k-1:k})=\F_k(r_k)$, $\F_k^2(\r_{k-1:k})={\bf 0}$,
$\H_k^1(\r_{k-1:k})=\H_k(r_k)$, $\H_k^2(\r_{k-1:k})={\bf 0}$, 
${\bf \Sigma}_k^{11}(\r_{k-1:k})$ $=\Q_k(r_k)$,
${{\bf \Sigma}_k^{21}(\r_{k-1:k})}^{T}=\H_k(r_k)\Q_k(r_k)$ 
and ${\bf \Sigma}_k^{22}(\r_{k-1:k})=\R_k(r_k)+\H_k(r_k) \Q_k(r_k)\H_k(r_k)^T$
corresponds to the linear and Gaussian JMSS 
\eqref{loi-jmss}.
Among models which satisfy \eqref{loi-pmc-conditional}-\eqref{condi-pmc-gaussien},
we now identify those which satisfy the following constraints:
%%
%%Remember that when the jumps are fixed, these models reduce
%%to PMC models of paragraph \ref{recall-pmc},
%%so independence properties of classical JMSS
%%recalled in Introduction are not necessarily 
%%satisfied by the conditional PMC models.
%%Now, we extend the classical linear
%%and Gaussian JMSS in the sense that we derive 
%%a class of conditional PMC models
%%in which the physical properties 
%%$p^1(r_k|r_{k-1})$, $f_{k|k-1}(\x_k|\x_{k-1},r_k)$ 
%%and $g_k(\y_k|\x_k)$ are kept.
%%%%%of a given 
%%%%%linear and Gaussian JMSS are kept, but not necessarily 
%%%%%the statistical ones.
%%So let a class of conditional linear and Gaussian
%%PMC models described by 
%%$p^{2,\theta}(\z_{0:k},\r_{0:k})$ such that:
\begin{itemize}
\item $\{r_k\}_{k\geq 0}$ is an MC with transitions 
$p^{2,\theta}(r_k|r_{k-1})=p^1(r_k|r_{k-1})$;
\item Given $\r_{0:k}$, $p^{2,\theta}(\x_k|\x_{k-1},\r_{0:k})=f_{k|k-1}(\x_k|\x_{k-1},r_k)$
and $p^{2,\theta}(\y_k|\x_{k},\r_{0:k})=g_{k}(\y_k|\x_{k},r_k)$.
\end{itemize}
By adapting the proof of Proposition \ref{prop-1},
we have the following proposition.
\begin{proposition}
\label{prop-2-saut}
Let us consider a set of pdfs of interest 
$p^1(\x_0|r_0)=\mathcal{N}(\x_0;\m_{0}(r_0);\P_{0}(r_0))$,
and $f_{k|k-1}(\x_k|\x_{k-1},r_k)$ and $g_{k}(\y_k|\x_k,r_k)$
given by \eqref{transition-jmss}-\eqref{vraisemblance-jmss}
%%%\begin{align}
%%%p^1(\x_0|r_0)&=\mathcal{N}(\x_0;\m_{0}(r_0);\P_{0}(r_0)) \text{,} \\
%%%f_{k|k-1}(\x_k|\x_{k-1},r_k)&=\mathcal{N}(\x_k;\F_k(r_k)\x_{k-1};\Q_k(r_k)) \text{ and } \\
%%%g_{k}(\y_k|\x_k,r_k)&=\mathcal{N}(\y_k;\H_k(r_k)\x_{k};\R_k(r_k)) \text{,}
%%%\end{align}
for all $k$.
Then the linear and Gaussian conditional PMC models which satisfy
\begin{eqnarray}
\label{condition-mc-saut} 
p^{2,\theta}(r_k|r_{k-1}) &=& p^1(r_k|r_{k-1}) \text{,} \\ 
\label{condition-1-saut} 
p^{2,\theta}(\x_k|\x_{k-1},\r_{0:k}) &=& f_{k|k-1}(\x_k|\x_{k-1},r_k) \text{,} 
\\
\label{condition-2-saut} 
p^{2,\theta}(\y_k|\x_{k},\r_{0:k}) &=& g_{k}(\y_k|\x_k,r_k) \text{,}
\end{eqnarray}
%%such that
%%$p_{k|k-1}(\x_k,\y_k|\x_{k-1},\y_{k-1},r_k,r_{k-1})$ does
%%not depend on parameters $(\m_0(r_0), \P_0(r_0))$
are described by the following 
equations:
\begin{align}
%%p^{2,\theta}(\z_0|r_0) \!\! &= \!\! \mathcal{N}\left(\z_0 ;\!\! \begin{bmatrix} \m_0(r_0) \\ \H_0(r_0)  \m_0(r_0)  \end{bmatrix} \!\!;\!\!
%%\begin{bmatrix}
%%\P_0(r_0)  & (\H_0(r_0)  \P_0(r_0) )^T \\
%%\H_0(r_0)  \P_0(r_0)  & \R_0(r_0)  \!\!+ \!\! \H_0(r_0)  \P_0(r_0)  \H_0(r_0) ^T 
%%\end{bmatrix} \right) \text{,}
%%\end{align}
p^{2,\theta}(r_k|r_{k-1}) &= p^1(r_k|r_{k-1}) \text{,} \\ 
p^{2,\theta}(\z_0|r_0)&=p^1(\x_0|r_0)g_0(\y_{0}|\x_0,r_0) \text{,} \\
%\end{align}
%%\begin{align}
%%\label{transition-prop-2}
%%p^{2,\theta}_{k|k-1}(\z_k|\z_{k-1},\r_{k-1:k}) \!\! &= \!\! \mathcal{N}(\z_k; \B_k(\r_{k-1:k}) \z_{k-1};
%%{\bf \Sigma_k}(\r_{k-1:k})) \text{,}
%%\end{align}
%\begin{align}
\label{transition-prop-2}
p^{2,\theta}_{k|k-1}(\z_k|\z_{k-1},\r_{k-1:k})  &=  \mathcal{N}(\z_k; \B_k(.) \z_{k-1};
{\bf \Sigma}_k(.)) \text{,}
\end{align}
where matrices $\B_k(\r_{k-1:k})$ and ${\bf \Sigma}_k(\r_{k-1:k})$
are defined by
\begin{align}
\label{matrice-B-saut}
& \B_k(\r_{k-1:k})=
%\nonumber \\ &
\begin{bmatrix}
\F_{k}(r_k)-\F_{k}^2(\r_{k-1:k})\H_{k-1}(r_{k-1}) & \F_{k}^2(\r_{k-1:k}) \\
\H_{k}(r_k)\F_{k}(r_k)-\H^2_{k}(\r_{k-1:k})\H_{k-1}(r_{k-1}) & \H^2_{k}(\r_{k-1:k})
\end{bmatrix} \text{,} \\
%{\bf \Sigma_k}&=
%\begin{bmatrix}
%\Q_{k} \!- \! \F_{k}^2 \R_{k-1} {\F_{k}^2}^T & [{\H}_{k}{\Q}_{k} \!- \!\H^2_{k}{\R}_{k-1}\F^2_{k}]^T \\
%{\H}_{k}{\Q}_{k} \!- \! \H^2_{k}{\R}_{k-1}{\F^2_{k}}^T &  \R_{k}\!-\! \H^2_{k}\R_{k-1}{\H^2_k}^T \! + \! \H_{k}\Q_{k}\H_{k} 
%\end{bmatrix} \text{,}
& {\bf \Sigma}_k(\r_{k-1:k})= \begin{bmatrix} {\bf \Sigma}_k^{11}(\r_{k-1:k}) & {{\bf \Sigma}_k^{21}(\r_{k-1:k})}^{T} \\ {\bf \Sigma}_k^{21}(\r_{k-1:k}) & {\bf \Sigma}_k^{22}(\r_{k-1:k}) \end{bmatrix} \text{,}
\end{align}
\begin{align}
\label{sigma-11-saut}
 {\bf \Sigma}_k^{11}(\r_{k-1:k}) &= \Q_{k}(r_k) -  \F_{k}^2(\r_{k-1:k}) \R_{k-1}(r_{k-1}) {\F_{k}^2(\r_{k-1:k})}^T \text{,} \\
\label{sigma-21-saut}
 {\bf \Sigma}_k^{21}(\r_{k-1:k}) & = {\H}_{k}(r_k){\Q}_{k}(r_k)
%% \nonumber \\ &
 - \H^2_{k}(\r_{k-1:k}){\R}_{k-1}(r_{k-1}){\F^2_{k}(\r_{k-1:k})}^T \text{,} \\
\label{sigma-22-saut}
 {\bf \Sigma}_k^{22}(\r_{k-1:k}) & =\R_{k}(r_k)\!-\! \H^2_{k}(\r_{k-1:k})\R_{k-1}(r_{k-1}){\H^2_k(\r_{k-1:k})}^T \!
%% \nonumber \\ &
  + \! \H_{k}(r_k)\Q_{k}(r_k)\H_{k}(r_k)^T \text{,} 
\end{align}
and where $\F_{k}^2(\r_{k-1:k})$ and $\H^2_{k}(\r_{k-1:k})$ can be arbitrarily chosen,
provided
%$\Q_{k}- \F_{k}^2 \R_k {\F_{k}^2}^T$, 
%$\R_{k}-\H^2_{k}\R_k{\H^2}^T+\H_{k}\Q_{k}\H_{k}$ and 
${\bf \Sigma}_k(\r_{k-1:k})$ is a positive definite covariance matrix
for all $k$.
\end{proposition} 
%%The proof is an adaptation to that of Proposition \ref{prop-1}.
%%%\begin{remark}
%%%\label{remark-5}
%%%{\rm
%%%In the same way that linear and Gaussian JMSS reduce to 
%%%linear and Gaussian HMC when discrete process $\r_{0:k}$ is
%%%known, the Conditional linear and Gaussian models derived 
%%%above reduce to those of Proposition \ref{prop-1} when 
%%%$\r_{0:k}$ is known. Remark that the probability distribution
%%%of $\r_{0:k}$ is identical in a given JMSS and models 
%%%derived above, so their comparison will actually rely on
%%%that of their respective underlying models, i.e. 
%%%the linear and Gaussian HMC model \eqref{transition-hmc}-\eqref{vraisemblance-hmc}
%%%and models derived in Proposition \ref{prop-1}. 
%%%}\end{remark}
\begin{remark}
\label{remark-jmss-tmc}
{\rm 
Of course, the particular setting $\F_k^2(\r_{k-1:k})={\bf 0}$,
$\H_k^2(\r_{k-1:k})={\bf 0}$ coincides
with the linear and Gaussian JMSS model \eqref{loi-jmss} which satisfies
\eqref{transition-jmss}-\eqref{vraisemblance-jmss}.
}
\end{remark}
The invariance properties of the models 
of Proposition \ref{prop-2-saut} are illustrated with the following
proposition and extend those of Propositions \ref{remark-1},
\ref{remark-2} and \ref{remark-3}.

\begin{proposition}
\label{proposition-invariance-saut}
In models of Proposition \ref{prop-2-saut},
$p^{2,\theta}(\z_k|\x_{k-1},$ $\r_{k-1:k})=p^1(\z_k|\x_{k-1},$ $\r_{k})$
and pdf $p^{2,\theta}(\x_{0:k},\r_{0:k})$ does not depend
on $\theta$: for all $\theta$,
\begin{align}
& p^{2,\theta}(\x_{0:k},\r_{0:k}) = p^1(\x_{0:k},\r_{0:k})= 
%\nonumber \\ & 
p^1(r_0)\prod_{i=1}^k p^1(r_i|r_{i-1}) p^1(\x_0|r_{0}) \prod_{i=1}^k f_{i|i-1}(\x_i|\x_{i-1},r_i) \text{;}
\end{align}
The difference with classical JMSS provides from pdf $p^{2,\theta}(\y_{0:k}|\x_{0:k},\r_{0:k})$
which now reads 
\begin{align}
& p^{2,\theta}(\y_{0:k}|\x_{0:k},\r_{0:k}) = p^{2,\theta}(\y_{0}|\x_{0:k},\r_{0:k})
%\times \nonumber \\ &
\prod_{i=1}^k p^{2,\theta}(\y_i|\x_{i-1:k},\y_{i-1}, \r_{i-1:k}) \text{.}
\end{align}
\end{proposition}

%%%\begin{remark}
%%%{\rm
%%%As in Remark \ref{remark-3}, $p^2(\x_{0:k}|\r_{0:k})$ 
%%%corresponds to $p^1(\x_{0:k}|\r_{0:k})$ 
%%%but the form of $p^2(\y_{0:k}|\x_{0:k},\r_{0:k})$ is more complicated
%%%than $p^1(\y_{0:k}|\x_{0:k},\r_{0:k})=\prod_{i=0}^k g_i(\y_i|\x_i,r_i)$: properties 
%%%\textbf{P.1} and \textbf{P.2} of classical JMSS
%%%are still satisfied but no longer \textbf{P.3}.
%%%}
%%%\end{remark}

\subsection{Exact Filtering in a subclass of constrained conditional linear and Gaussian PMC models}
\label{exact-filtering}

\subsubsection{Preliminary result}
The problem we address now is the computation
of $\Phi_k$ in the class of constrained
conditional linear and Gaussian PMC models
of Proposition \ref{prop-2-saut}.
Of course, the exact computation of
$\Phi_k$ in \eqref{Theta_n-def-generale}
is not possible in all models of
Proposition \ref{prop-2-saut};
otherwise, it would be also possible 
in the linear and Gaussian JMSS $p^1(\z_{0:k},\r_{0:k})$
since $p^1(.)$ is a particular model out of this class
(see Remark \ref{remark-jmss-tmc}). 
%%For this general class of models, 
%%PF can be adapted
%%\cite{nsip2003} \cite{abassi-triplet}.
%%Indeed, $p^2(\x_{k}|\y_{0:k},\r_{0:k})$ 
%%is computable via the KF for PMC models 
%%\cite{KalmanPairwise}
%%while $p^2(\r_{0:k})$ can be approximated 
%%via PF. We do not detail the extension of 
%%the PF for this class of models.
However, we show that for a particular setting
of $\H_k^2(\r_{k-1:k})$ in \eqref{matrice-B-saut}, 
the computation of $\Phi_k$ at a linear 
computational cost becomes possible. 
%%For clarity, we set $f(\x)=\x$ in \eqref{Theta_n-def-generale}.
In a general conditional linear and
Gaussian PMC model \eqref{loi-pmc-conditional}-\eqref{condi-pmc-gaussien},
we have this preliminary result (a proof is given in Appendix \ref{proof-exact-filtering})
when given $\z_{k-1}$ and $\r_{k-1:k}$, $\y_{k}$ 
does not depend on $\x_{k-1}$, i.e. when 
$p(\y_k|\z_{k-1},\r_{k-1:k})=$ $p(\y_k|\y_{k-1},\r_{k-1:k})$:
\begin{proposition}
\label{proposition-filtering}
Let a conditional linear and Gaussian PMC model which satisfies 
\begin{align}
\label{condition-filtering}
 p^2(\z_k|\z_{k-1},\r_{k-1:k})&= p^2(\y_k|\y_{k-1},\r_{k-1:k})
% \times \nonumber \\ &
 p^2(\x_k|\z_{k-1},\y_k,\r_{k-1:k}) \text{,} \\
\label{loi-conditionnelle-saut-1}
 p^2(\x_k|\z_{k-1},\y_{k},\r_{k-1:k})&=\mathcal{N}(\x_k;{\bf C}_{k}(\r_{k-1:k})\x_{k-1} +
%\nonumber \\ &
{\bf D}_{k}(\r_{k-1:k},\y_{k-1:k}); {\bf \Sigma}_{k}^\x(\r_{k-1:k})) \text{.} 
\end{align}
Then $p^2(r_{k}|\y_{0:k})$,
${\rm E}({\bf x}_{k}|\y_{0:k},r_{k})$
and
${\rm E}({\bf x}_{k}\x_k^T|\y_{0:k},r_{k})$
can be computed recursively via
\begin{align}
\label{proba-r-filtering}
p^2(r_{k}|\y_{0:k}) &\propto 
\sum_{r_{k-1}} p^2(r_{k}|r_{k-1})p^2(\y_{k}|\y_{k-1},\r_{k-1:k})
%\times \nonumber \\ &
p^2(r_{k-1}|\y_{0:k-1}) \text{,} \\
\label{exact-filtering-esperance}
{\rm E}({\bf x}_{k}|\y_{0:k},r_{k}) \! &= \!
\sum_{r_{k-1}} \! p^2(r_{k-1}|r_{k},\y_{0:k}) ( {\bf C}_{k}(\r_{k-1:k})
%\times \nonumber \\ &
{\rm E}({\bf x}_{k-1}|\y_{0:k-1},r_{k-1}) \!+ \! {\bf D}_{k}(\r_{k-1:k},\y_{k-1:k}) ) \text{,} \\
%%\end{align}
%%
%%\begin{align}
\label{exact-filtering-variance}
{\rm E}(\x_{k}\x_k^T|\y_{0:k},r_{k}) \! &= \!
\sum_{r_{k-1}} \! p^2(r_{k-1}|r_{k},\y_{0:k}) \times
\big( {\bf \Sigma}_{k}^\x(\r_{k-1:k}) 
\nonumber \\ &
+ {\bf C}_{k}(\r_{k-1:k}) {\rm E}({\bf x}_{k-1}{\bf x}_{k-1}^T|\y_{0:k-1},r_{k-1}) {\bf C}_{k}(\r_{k-1:k})^T \! \! 
\nonumber \\ &
+
{\bf D}_{k}(\r_{k-1:k},\y_{k-1:k}) ({\rm E}({\bf x}_{k-1}|\y_{0:k-1},r_{k-1}))^T{\bf C}_{k}(\r_{k-1:k})^T 
\nonumber \\ 
&
+
{\bf C}_{k}(\r_{k-1:k}){\rm E}({\bf x}_{k-1}|\y_{0:k-1},r_{k-1}){\bf D}_{k}(\r_{k-1:k},\y_{k-1:k})^T 
\nonumber \\ &
+
{\bf D}_{k}(\r_{k-1:k},\y_{k-1:k}){\bf D}_{k}(\r_{k-1:k},\y_{k-1:k})^T \big) \text{,} \\ 
%\end{align}
%\begin{align}
\label{proba-r-filtering-2}
p^2(r_{k-1}|r_{k},\y_{0:k}) & \propto p^2(r_{k}|r_{k-1})p^2(\y_{k}|\y_{k-1},\r_{k-1:k})
%\times \nonumber \\ &
p^2(r_{k-1}|\y_{0:k-1}) \text{.} 
\end{align}
\end{proposition}
The computation of $\Phi_k$ 
is deduced from 
$\Phi_k=$ $ \sum_{r_{k}} p^2(r_{k}|\y_{0:k})$ $ {\rm E}(f({\bf x}_{k})|\y_{0:k},r_{k})$.
Remark that the computational cost involved in the computation of
$\Phi_k$ is no longer exponential, but is indeed linear in time
and only requires sums on $r_{k-1}$ and $r_k$, at time $k$.

\subsubsection{Application to the physically constrained PMC models with jumps}
Now we turn back to the class of models of Proposition \ref{prop-2-saut}
and we look for those which satisfy constraint \eqref{condition-filtering}.
In this class of models, $p^{2,\theta}(\y_k|\z_{k-1},\r_{k-1:k})$ depends
on $\x_{k-1}$ via its mean which reads
$
\left(\H_{k}(r_k)\F_{k}(r_k)-\H^2_{k}(\r_{k-1:k})\H_{k-1}(r_{k-1})\right)\x_{k-1} + \H^2_{k}(\r_{k-1:k}) \y_{k-1} \text{;}
$
so $p^{2,\theta}(\y_k|\z_{k-1},\r_{k-1:k})$ does not depend on $\x_{k-1}$
(and so the exact computation of $\Phi_k$ is possible according to
Proposition \ref{proposition-filtering}) 
if one can find $\H^2_{k}(\r_{k-1:k})$ which satisfies
\begin{align}
\label{condition-H-2}
\H_{k}(r_k)\F_{k}(r_k)-\H^2_{k}(\r_{k-1:k})\H_{k-1}(r_{k-1})={\bf 0} \text{.}
\end{align}
The expression of $p^{2,\theta}(\x_k|\z_{k-1},$ $\y_k,$ $\r_{k-1:k})$
is deduced from that of $p^{2,\theta}(\z_k|\z_{k-1},\r_{k-1:k})$
in \eqref{transition-prop-2} 
(see Lemma \ref{lemme-2} in Appendix \ref{appendice-lemma}),
and we have the following corollary.
\begin{corollary}
Let a constrained conditional linear and Gaussian PMC model
of Proposition \ref{prop-2-saut}
which satisfies \eqref{condition-H-2}.
Then the exact computation of $\Phi_k$ 
is possible by using Proposition \ref{proposition-filtering}
with 
\begin{align}
\label{matrice-C}
 {\bf C}_{k}(r_{k-1:k})& =  \F_{k}(r_k)  -  \F_{k}^2(\r_{k-1:k}) \H_{k-1}(r_{k-1}) \text{,} \\
\label{matrice-D}
 {\bf D}_{k}(\r_{k-1:k},\y_{k-1:k})&=\F_{k}^2(\r_{k-1:k}) \y_k +
({\bf \Sigma}_k^{21}(\r_{k-1:k}))^T \times
%\nonumber \\ &
({\bf \Sigma}_k^{22}(\r_{k-1:k}))^{-1}
(\y_{k}-\H^2_{k}(\r_{k-1:k})\y_{k-1}) \text{,} \\
\label{matrice-sigma-x}
{\bf \Sigma}_{k}^\x(\r_{k-1:k})&= {\bf \Sigma}_k^{11}(\r_{k-1:k}) -
({\bf \Sigma}_k^{21}(\r_{k-1:k}))^T
%\times \nonumber \\ &
({\bf \Sigma}_k^{22}(\r_{k-1:k}))^{-1}{\bf \Sigma}_k^{21}(\r_{k-1:k}) \text{,} \\
 p^2(\y_k|\y_{k-1},\r_{k-1:k})&= \mathcal{N}(\y_k;\H_k^2(\r_{k-1:k})\y_{k-1};{\bf \Sigma}_k^{22}(\r_{k-1:k})) \text{.} 
\end{align}
\end{corollary} 

\subsubsection{Summary}
Let us summarize the discussion. Thus far, we
have proposed a class of stochastic models 
$p^{2,\theta}(\z_{0:k},\r_{0:k})$ which satisfy
the constraints $p^{2,\theta}(r_k|r_{k-1})=p^1(r_k|r_{k-1})$,
$p^{2,\theta}(\x_k|\x_{k-1},r_k)=f_{k|k-1}(\x_k|\x_{k-1},r_k)$ 
and $p^{2,\theta}(\y_k|\x_{k},r_k)=g_k(\y_k|\x_k,r_k)$
and in which $\Phi_k$ can be computed exactly (no Monte Carlo nor 
numerical approximations are needed) and at a computational
cost which is linear in the number of observations. 
This algorithm can be applied whenever a problem is
essentially described by the physical linear 
and Gaussian properties $f_{k|k-1}(\x_k|\x_{k-1},r_k)$ and
$g_k(\y_k|\x_k,r_k)$.
%%for approximately
%%computing $\Phi_k$ in a linear and Gaussian JMSS.
%%So we are now able to propose a new algorithm
%%to compute $\Phi_k$ in a linear and Gaussian JMSS.
%%%\begin{flushleft}
%%%{\textbf{A new approximate algorithm in linear and Gaussian JMSS}}
%%%\end{flushleft}
So let physical properties %a linear and Gaussian JMSS 
\eqref{transition-jmss}-\eqref{vraisemblance-jmss}
parametrized by $\F_k(r_k)$, $\H_k(r_k)$, $\Q_k(r_k)$
and $\R_k(r_k)$. The goal is to compute 
${\rm E} (f(\x_k)|\y_{0:k})$ recursively via 
$ p^{2,\theta}(r_k|\y_{0:k})$ and ${\rm E} (f(\x_k)|\y_{0:k},r_k)$.
The algorithm is as follows. At time $k-1$, we have
$p(r_{k-1}|\y_{0:k-1})$,
${\rm E}({\bf x}_{k-1}|\y_{0:k-1},r_{k-1})$
and  ${\rm E}({\bf x}_{k-1}{\bf x}_{k-1}^T|\y_{0:k-1},r_{k-1})$;
for $r_{k-1:k} \in \{1,\cdots,K\} \times \{1,\cdots,K\} $,
\begin{itemize}
\item[{\bf S.1}]
Deduce the class of linear and Gaussian PMC models 
parametrized by $\F_k^2(\r_{k-1:k})$,
$\H_k^2(\r_{k-1:k})$ using Proposition \ref{prop-2-saut};
\item[{\bf S.2}] Choose $\H_k^2(\r_{k-1:k})$ such that $\H_{k}(r_k)\F_{k}(r_k)-\H^2_{k}(\r_{k-1:k})\H_{k-1}(r_{k-1})={\bf 0}$; 
\item[{\bf S.3}] Compute matrices ${\bf C}_{k}(r_{k-1:k})$,
${\bf D}_{k}(r_{k-1:k})$ and ${\bf \Sigma}_{k}^\x(\r_{k-1:k})$ using
\eqref{matrice-C}-\eqref{matrice-sigma-x};
\item[{\bf S.4}]
%For $r_k \in \{1,\cdots,K \}$,
Compute $p^{2,\theta}(r_{k}|\y_{0:k})$,
${\rm E}({\bf x}_{k}|\y_{0:k},r_{k})$
and ${\rm E}({\bf x}_{k}{\bf x}_{k}^T|\y_{0:k},r_{k})$
via \eqref{proba-r-filtering}-\eqref{proba-r-filtering-2}.
%and deduce $\Phi_k$ via \eqref{esperance-x}.
\end{itemize}
Finally, compute ${\rm E} (f(\x_k)|\y_{0:k})=$ $\sum_{r_k} p^{2,\theta}(r_k|\y_{0:k})$  ${\rm E} (f(\x_k)|\y_{0:k},r_k)$.

\subsection{A particular application: approximate $\Phi_k$ in a linear and Gaussian JMMS}
Until now, we have proposed a class of exact filtering algorithms
when the problem involves given physical properties
of interest. Now, remember that the linear and Gaussian JMSS
$p^1(.)$ shares those physical properties with the class of models
$p^{2,\theta}(.)$ in which 
the optimal Bayesian estimate can be computed. So the approximation
of $\Phi_k$ in a linear and Gaussian JMSS via our exact filtering
algorithm arises naturally at this point. So in this section let us assume
that the data indeed follow a linear and Gaussian JMSS 
\eqref{loi-jmss}-\eqref{vraisemblance-jmss}.
Since our algorithm is parametrized by $\F_k^2(\r_{k-1:k})$,
%%%We have just described a generic algorithm for approximating 
%%%$\Phi_k$ in a given JMSS \eqref{loi-jmss}-\eqref{vraisemblance-jmss}.
%%%However this algorithm is parametrized by matrices $\F_k^2(\r_{k-1:k})$,
it remains to choose $\F_k^2(\r_{k-1:k})$ which best fits the
original model.

%%%\begin{remark} 
%%%{\rm
%%%Although our first objective 
%%%was to derive an alternative 
%%%filtering technique in linear and Gaussian JMSS,
%%%it appears that our solution can be used for the general class of
%%%models of the Proposition \ref{prop-2-saut}, which includes
%%%the classical linear and Gaussian JMSS but also models 
%%%which have more general statistical properties.
%%%Therefore, parameters $\F_k^2(\r_{k-1:k})$
%%%can be adjusted in function of the data, which
%%%may follow a more complicated model than
%%%the classical JMSS one. This case does not
%%%prevent from the computation of $\Phi_k$,
%%%as long as \eqref{condition-H-2} is satisfied. 
%%%}
%%%\end{remark}
%%Now we shall discuss on the choice of $\F_k^2(\r_{k-1:k})$
%%when data indeed follow the given linear and Gaussian
%%JMSS model \eqref{transition-jmss}-\eqref{vraisemblance-jmss}.
In a linear and Gaussian JMSS, $\F_k^2(\r_{k-1:k})= {\bf 0}$
and $\H_k^2(\r_{k-1:k})= {\bf 0}$.
However, $\F_k^2(\r_{k-1:k})= {\bf 0}$ should not be our choice here, as we now see, 
because in our models, $\H_k^2(\r_{k-1:k})$ is different of  ${\bf 0}$ 
from constraint \eqref{condition-H-2}.
The idea is to tune $\F_k^2(\r_{k-1:k})$
such that constraint \eqref{condition-H-2} is balanced. 
More precisely, we look for $\F_k^2(\r_{k-1:k})$ such that the Kullback-Leibler
Divergence (KLD) between $p^{2,\theta}(\z_{0:k},\r_{0:k})$, which satisfies
\eqref{condition-H-2}, and $p^{1}(\z_{0:k},\r_{0:k})$ is minimum.
We have the following result (a proof is given in Appendix \ref{proof-prop-4}).%%%
\begin{proposition}
\label{proposition-KL}
Let $p^1(.)$ be the linear and Gaussian JMSS model
and $p^{2,\theta}(.)$ be the class of models 
of Proposition \ref{prop-2-saut} which satisfy condition \eqref{condition-H-2}.
Parameters $\F_k^{2}(\r_{k-1:k})$ which minimize the KLD
between  $p^{2,\theta}(\z_{0:k},\r_{0:k})$ and $p^1(\z_{0:k},\r_{0:k})$
are given by
\begin{align}
\label{condition-F-2}
& \F_k^{2,{\rm opt}}(\r_{k-1:k})= \Q_k(r_k) \H_k(r_k)^T
%\times \nonumber \\ &
\left [\R_k(r_k) + \H_k(r_k) \Q_k(r_k) \H_k(r_k)^T \right]^{-1} \H_k^2(\r_{k-1:k}) \text{.}
\end{align}
\end{proposition}
Remember however that these particular parameters should be used when the goal
is to approximate the computation of the optimal estimator $\Phi_k$
in \eqref{Theta_n-def-generale}
in a linear and Gaussian JMSS. For more general models (for which our
filtering technique can be still used), these parameters do not guarantee the
best performances, as we will see in the Simulations section.

%%This choice minimizes the KLD between pdf $p^1(\z_{0:k},\r_{0:k})$
%%of classical linear and Gaussian JMSS and pdf $p^2(\z_{0:k},\r_{0:k})$ of models
%%of the Proposition \ref{prop-2-saut} which satisfy \eqref{condition-H-2-pmc}. 
%%Indeed, 
%%and for a given $\r_{0:k}$, $ {\rm D}_{\rm KL}(p^2(\z_{0:k}|\r_{0:k}),p^1(\z_{0:k}|\r_{0:k}))$ 
%%is minimum according to Remark \ref{remark-DKL}.

\section{Performance Analysis and Simulations}
\label{section-simulation}

In this section, we start by describing step by step
our methodology and we validate our discussions: 
we first generate data from a given HMC model and
we estimate the hidden data with a filter based on a PMC model
out of the class described by Proposition \ref{prop-1}
which satisfies conditions \eqref{condition-H-2} and
\eqref{condition-F-2} when jumps are fixed.
We compare the performance
of this approximation with the optimal KF.
Next, we compare our new approximate filtering solution
for linear and Gaussian JMSS with the IMM algorithm and 
the PF. When simulations are involved, 
we generate, for a given model,
$P=200$ sets of data of length $T=100$.

\subsection{A step by step illustration}
Let us describe our methodology 
step by step on the popular scalar model with jumps,
%(${\rm dim}(\x)={\rm dim}(\y)=1$)
($p=m=1$)
(see e.g. \cite{fernhead-jmss}\cite{Chen-Mixture-Kalman} and references therein): 
\begin{align}
\label{simu-saut-hmc-1}
f_{k|k-1}(x_k|x_{k-1},r_k)&=\mathcal{N}(x_k;a(r_k)x_{k-1};Q(r_k)) \text{,} \\
\label{simu-saut-hmc-2}
g_{k}(y_k|x_{k},r_k)&=\mathcal{N}(x_k;b(r_k)x_{k};R(r_k)) \text{,}
\end{align}
where $|a(r_k)| \leq 1$ and $\{r_k \}_{k \geq 0}$ is a given MC with
transition probabilities $p^1(r_k|r_{k-1})$.
First, we omit the jumps and we consider 
the underlying model 
described by the two following pdfs:
\begin{align}
\label{simu-hmc-1}
f_{k|k-1}(x_k|x_{k-1})&=\mathcal{N}(x_k;ax_{k-1};Q) \text{,} \\
\label{simu-hmc-2}
g_{k}(y_k|x_{k})&=\mathcal{N}(y_k;bx_{k};R) \text{,}
\end{align}
where $|a| \leq 1$.
Next, remember that we need to deduce the associated
class of linear and Gaussian PMC models which satisfy the same
physical properties \eqref{simu-hmc-1}-\eqref{simu-hmc-2}.
They are described by two parameters $F_k^2=c$ and 
$H_k^2=d$, which gives a class of PMC models described by the following transition according to
Proposition \ref{prop-1}:
\begin{align}
\label{pmc-exemple-1}
p^{2,\theta}(\z_k|\z_{k-1})&= \mathcal{N}\big(\z_k;
\begin{bmatrix}
a-bc & c \\
ab-db & d
\end{bmatrix}\z_{k-1};
%\nonumber \\&
\begin{bmatrix}
Q-c^2R & bQ-cdR \\
bQ-cdR & R(1-d^2)+b^2Q
\end{bmatrix} \big) \text{.}
\end{align}
According to \eqref{condition-H-2},
we look for parameter $d$ such that
$ab-db =0$, so from now on we set $d=a$.

Next, if the goal is to approximate the HMC model
deduced from \eqref{simu-hmc-1}-\eqref{simu-hmc-2}, 
the parameter $c$ which minimizes the KLD
between $p_{k|k-1}^{2,\theta}(\z_{k}|\z_{k-1})$ and $p_{k|k-1}^1(\z_{k}|\z_{k-1})$,
is $c=\frac{abQ}{R+b^2Q}$, from \eqref{condition-F-2}; so among all PMC 
models \eqref{pmc-exemple-1}, we choose 
\begin{align}
\label{pmc-exemple-optimal}
p^{2,\theta}_{k|k-1}(\z_k|&\z_{k-1})= \mathcal{N} \big(\z_k;
\begin{bmatrix}
a-\frac{ab^2Q}{R+b^2Q} & \frac{abQ}{R+b^2Q} \\
0 & a
\end{bmatrix} \z_{k-1};
%\nonumber \\&
\begin{bmatrix}
Q-\frac{ a^2b^2Q^2R}{(R+b^2Q)^2} & bQ-\frac{a^2bQR}{R+b^2Q} \\
bQ-\frac{a^2bQR}{R+b^2Q} & R(1-a^2)+b^2Q
\end{bmatrix} \!\!
\big) \text{.}
\end{align}
It is easy to check that the covariance matrix 
of $p^{2,\theta}_{k|k-1}(\z_k|\z_{k-1})$ is
positive definite, whatever $-1 \leq a \leq 1$, $b$, $Q$ 
and $R$.
It is now interesting to compare the KLD
between $p_{k|k-1}^{2,\theta}$ and $p_{k|k-1}^1$ which reduces
to that between $p^{2,\theta}(y_k|y_{k-1})$ and $p^1(y_k|x_{k-1})$
since we have chosen the optimal parameter $c$
(see the proof of Proposition \ref{proposition-KL}).
Remember that in HMC \eqref{simu-hmc-1}-\eqref{simu-hmc-2}, 
$p^1(y_k|x_{k-1})=\mathcal{N}(y_k;abx_{k-1};b^2Q+R)$
and in PMC \eqref{pmc-exemple-optimal},
$p^{2,\theta}(y_k|y_{k-1})=\mathcal{N}(y_k;ay_{k-1};R(1-a^2)+b^2Q)$;
using classical results on the KLD between two Gaussians (see e.g. \cite{ref-dkl}),
we have
\begin{align}
&{\rm D}_{\rm KL}(p^{2,\theta}(y_k|y_{k-1}),p^1(y_k|x_{k-1}))= 0.5 \times
%\nonumber \\ &
\left[-\frac{a^2R}{R+b^2Q} \! + \! \frac{a^2(y_{k-1}-bx_{k-1})^2}{R+b^2Q} \!- \! {\rm ln} (\frac{R+b^2Q-a^2R}{R+b^2Q})\right] \text{,} 
\end{align}
which depends on r.v $y_{k-1}$ and $x_{k-1}$ via
$(y_{k-1}-bx_{k-1})^2$. However, in such models 
${\rm E}((y_{k-1}-bx_{k-1})^2)=R$, so 
\begin{align}
\label{mean-KLD-exemple}
{\rm E}({\rm D}_{\rm KL}&(p^{2,\theta}(y_k|y_{k-1}),p^1(y_k|x_{k-1}))) =
%%&= -0.5{\rm ln} (1-\frac{a^2R}{R+b^2Q}) \text{,} \\
%\nonumber \\ & 
%\label{mean-KLD-exemple-2}
-0.5{\rm ln} (1-\frac{a^2(R/Q)}{R/Q+b^2}) \text{.}
\end{align}
It is an increasing function of ratio $R/Q$, 
so when $R/Q$ is small,
i.e. the process noise is large compared to the
observation one, then PMC model \eqref{pmc-exemple-optimal}
is close to the original HMC model deduced from \eqref{simu-hmc-1}-\eqref{simu-hmc-2},
so estimating the hidden data from \eqref{pmc-exemple-optimal}
(although they follow \eqref{simu-hmc-1}-\eqref{simu-hmc-2}) should not
have a serious impact.

\begin{figure} [htbp!]
\center
	\includegraphics[width=85.7mm]{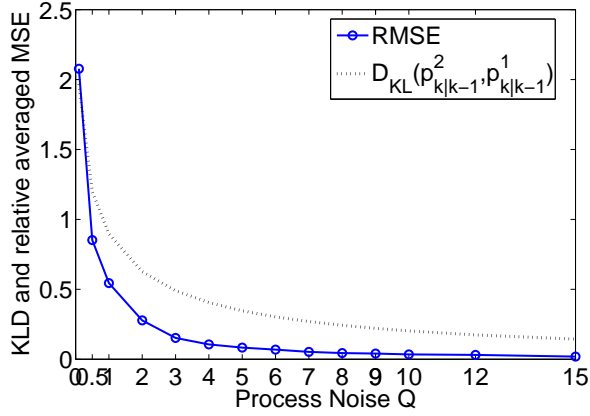} 
	\caption{RMSE between a classical KF based on \eqref{simu-saut-hmc-1}-\eqref{simu-saut-hmc-2} and a PMC-KF based on 
	\eqref{pmc-exemple-optimal} (blue circle) and KLD between transitions of the HMC model based on \eqref{simu-saut-hmc-1}-\eqref{simu-saut-hmc-2} and
model \eqref{pmc-exemple-optimal} (black dotted line). When $Q$ increases, both RMSE and DKL decrease;
the estimates based on
model \eqref{pmc-exemple-optimal} 
are very close to the optimal ones.}
\label{RMSE-vs-KLD}
\end{figure}

We generate data from the HMC
model deduced from \eqref{simu-hmc-1}-\eqref{simu-hmc-2}
where we set $a=b=R=1$.
We compute a KF for PMC \cite{KalmanPairwise}
based on model
%which verifies 
%\eqref{condition-H-2-pmc} and
%\eqref{condition-F-2}
\eqref{pmc-exemple-optimal}
and the KF for classical model
\eqref{simu-hmc-1}-\eqref{simu-hmc-2}, which of course
is optimal for this model in the sense that is minimizes the MSE.
We note $\hat{x}_{k,p,1}$ (resp. $\hat{x}_{k,p,2}$)
the estimator based on the original HMC model (resp.based 
on the PMC model) for the $p$-th simulation at time
$k$. For each estimate, we compute the averaged
MSE over time:
\begin{align}
\mathcal{J}^i&= \frac{1}{T} \sum_{k=1}^T \left[ \frac{1} {P} \sum_{p=1}^P
(\hat{x}_{k,p,i}-x_{k,p})^2 \right]
\end{align}
where $x_{k,p}$ is the true state
for the $p$-th realization at time $k$.
We also compute the mean of the KLD 
\eqref{mean-KLD-exemple} between 
$p_{k|k-1}^1$ and $p_{k|k-1}^{2,\theta}$. 
In Figure \ref{RMSE-vs-KLD}, we display the KLD
between $p_{k|k-1}^1$ and $p_{k|k-1}^{2,\theta}$
and the relative averaged MSE (RMSE) $(\mathcal{J}^1-\mathcal{J}^2)/\mathcal{J}^2$
against $Q$.
As expected, the RMSE decreases 
when ${\rm D}_{\rm KL}(p^{2,\theta}_{k|k-1},p^1_{k|k-1})$ decreases,
i.e. when $Q$ increases.
Particularly interesting, values of RMSE
are below $0.10$ when $Q \geq 4$ and for high values of $Q$ ($Q=10$),
they are close to $0.03$.
It means that approximating the original HMC model
with a PMC one of Proposition \ref{prop-1}
in which $H_k^2$ and $F_k^2$ respectively 
satisfy \eqref{condition-H-2} and
\eqref{condition-F-2} does not differ to the optimal
method as long as $Q$ is not too small.

\subsection{Performance Analysis on jumps Scenario} 
We now consider two scenarios with jumps.
We compute our solution of paragraph
\ref{exact-filtering} ($\hat{x}_{k,p,1}$),
a PF based on the SIR algorithm with the importance
distribution $p^1(r_k|r_{k-1})$ (it only requires
one KF per particle) with $N=100$ particles 
\cite{doucet-jump-Mkv} ($\hat{x}_{k,p,2}$),
an IMM algorithm \cite{Shalom-IMM} ($\hat{x}_{k,p,3}$)
and a KF ($\hat{x}_{k,p,{\rm Kalm}}$) which uses the 
true jumps and which is our benchmark solution. We compute
the MSE between the estimates and the estimator based
on the KF which uses the true jumps:
\begin{align}
{\rm MSE}^i(k)&=\frac{1}{P} \sum_{p=1}^P (\hat{x}_{k,p,i}-\hat{x}_{k,p,{\rm Kalm}})^2 \text{.}
\end{align}

\subsubsection{Scalar model with jumps}
\begin{figure} [htbp!]
\center
\begin{subfigure}[]
{
\centering
\includegraphics[height=0.23 \textheight]{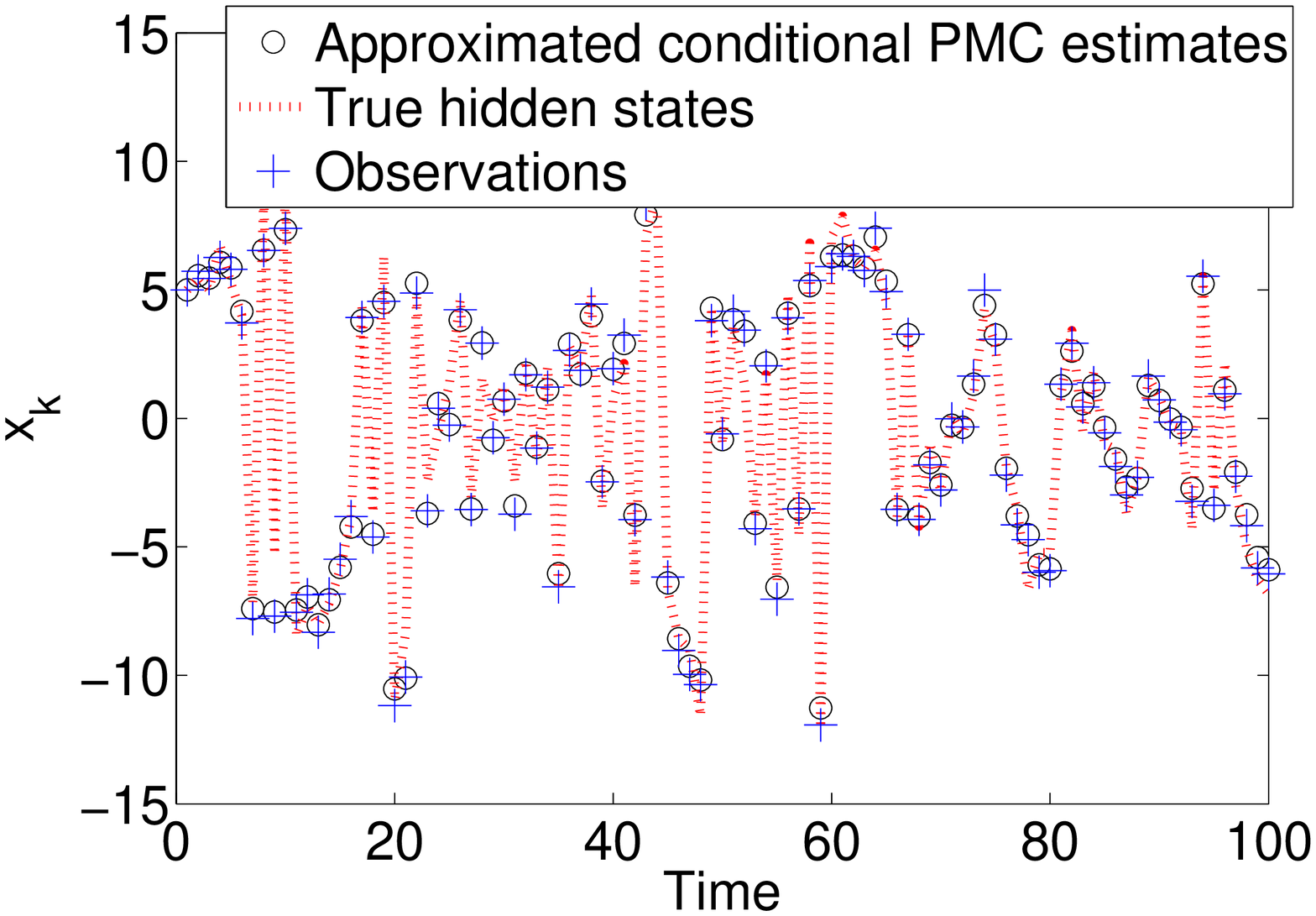} 
	\label{scenario-1D-saut}
}
\end{subfigure} 
\begin{subfigure} []
{
\centering
\includegraphics[height=0.23 \textheight]{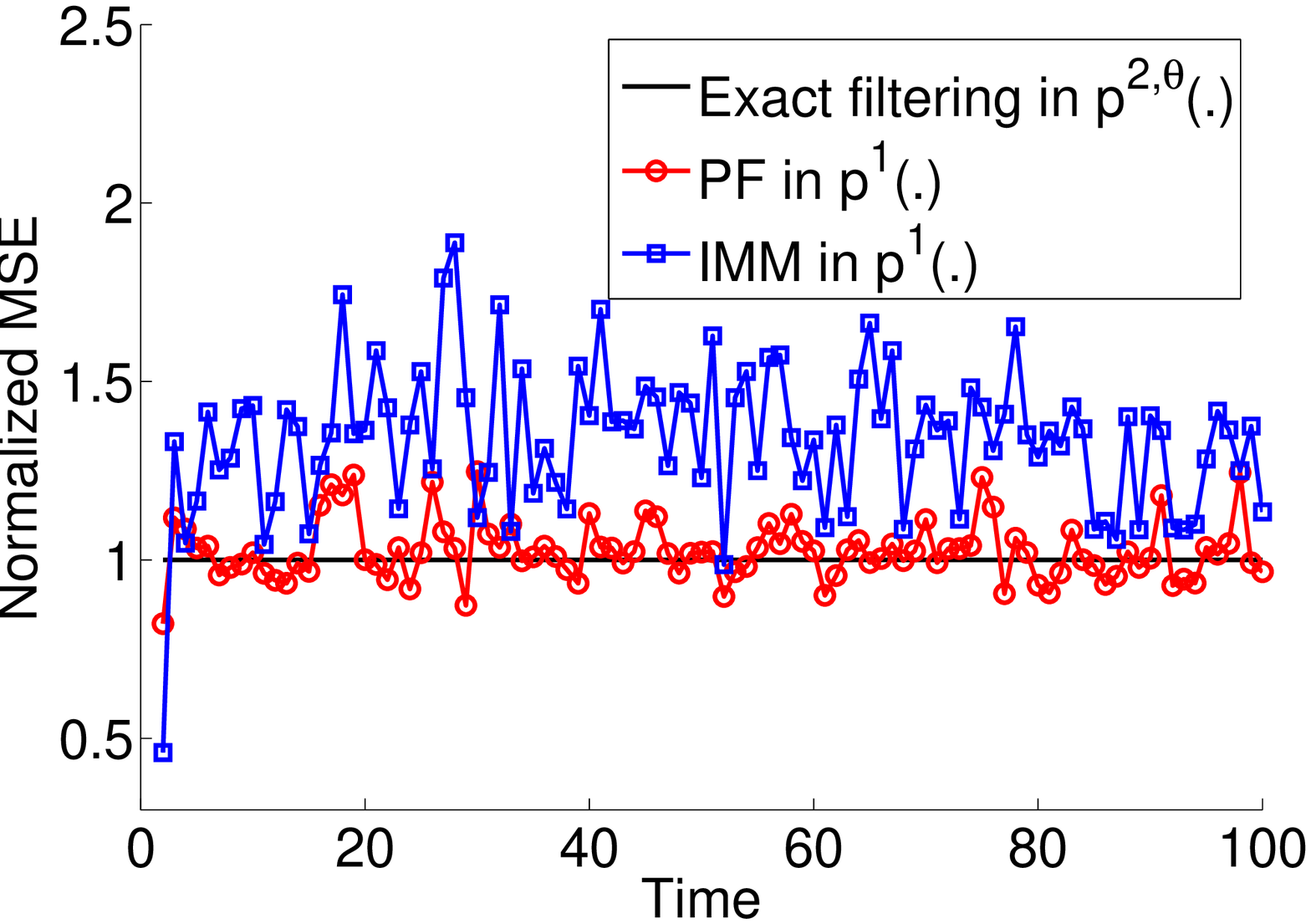} 
\label{modele-1D-saut-MSE}
}
\end{subfigure}
\caption{(a) - Example of scenario of model \eqref{simu-saut-hmc-1}-\eqref{simu-saut-hmc-2}
and restoration with a conditional PMC model of the Proposition \ref{prop-2-saut} which 
satisfies \eqref{condition-H-2} and \eqref{condition-F-2}.
True states (red dotted line), estimates based on our new approximation (black circles) and observations (blue crosses).
(b) - Normalized MSE of our algorithm (black line),
PF (red circles) and IMM (blue squares) estimates.}
\end{figure}

We go on with model 
\eqref{simu-saut-hmc-1}-\eqref{simu-saut-hmc-2}
where $r_k \in \{1,2,3 \}$, $a_k(r_k)=[1,-0.9,0.9]$, $b=1$,
$Q(r_k)=[3,10,10]$ and $R=1$. 
The transition probabilities are defined by
$p^1(r_{k}|r_{k-1})=0.8$ if $r_{k}=r_{k-1}$ and 
$p^1(r_{k}|r_{k-1})=0.1$ if $r_{k}\neq r_{k-1}$.
Data are generated from the JMSS \eqref{loi-jmss}.
A typical scenario is displayed in Fig. \ref{scenario-1D-saut}.
Remember from the previous paragraph that
our new approximation filtering technique is based on
the conditional linear and Gaussian PMC model 
\begin{align}
p_{k|k-1}^{2,\theta}(\z_k|\z_{k-1},\r_{k-1:k})\!\!&=\!\! \mathcal{N}(\z_k; 
\B_k(\r_{k-1:k})\z_{k-1}; {\bf \Sigma}(\r_{k-1:k})) \text{,} \\
\B_k(\r_{k-1:k}) \! &= \!
\begin{bmatrix}
a(r_k)-\frac{a(r_k) b^2 Q(r_k) }{R + b^2 Q(r_k) } & \frac{ a(r_k)b Q(r_k) }{R(r_k)+ b^2Q(r_k) } \\
0 & a(r_k)
\end{bmatrix} \text{,} \\
 {\bf \Sigma}(\r_{k-1:k})&=
%%\nonumber \\ &
\!\!
\begin{bmatrix}
Q(r_k)-\frac{a(r_k)^2 b^2 Q(r_k)^2 R}{(R+b^2 Q(r_k))^2} \!\! & \!\! bQ(r_k) -\frac{a(r_k)^2 b Q(r_k) R }{R+b^2Q(r_k)} \\
bQ(r_k)-\frac{a(r_k)^2 b Q(r_k) R }{R+b^2Q(r_k)} \!\! &  \!\! R(1-a(r_k)^2)+ b^2Q(r_k)
\end{bmatrix} \!\! \text{.}
\end{align}
MSEs of the different estimates
are displayed in Fig \ref{modele-1D-saut-MSE}
and are normalized w.r.t. that of our approximated solution. 
Particularly interesting,
we see that our algorithm outperforms the IMM estimate 
and slightly improves (in mean) the PF. However,
remember that our technique is not based on Monte Carlo samples
and is more interesting from a computational point of view.
In order to illustrate this difference, 
we have computed the ratio of the averaged computational 
time used by the PF and our solution which is approximately 
equal to $15$: our solution is thus much faster than SMC methods.

\begin{remark}
{\rm 
If we increase the number of particles, the performances of the 
PF are improved and are identical to those of our exact filtering technique. 
Thus, it may be interesting to average the efficiency
${\rm Eff}(k) = \frac{1}{ {\rm MSE}(k) {\rm E}(C(k))}$ over time where 
$C(k)$ is the CPU time to compute the estimate. The efficiency
of our algorithm does not depend on the number of particles and is $8.5 \times 10^4$
while for the PF the efficiency decreases when the number of particles
increases and varies between $5 \times 10^3$ for $100$ particles and $0.1\times 10^3$ for $1000$ particles.
%%%This can be seen from the 
%%%Figure XXX where we have computed the average of the efficiency in function
%%%of the number of particles for the three
%%%estimates, defined as ${\rm Eff}(k) = \frac{1}{ {\rm MSE}(k) {\rm E}(C(k))}$ where 
%%%$C(k)$ is the CPU time to compute the estimate. Of course, the IMM and our
%%%exact filtering algorithms do not depend on the number of particles
}
\end{remark}

\begin{figure} [htbp!]
\center
\end{figure}

\subsubsection{Target Tracking}
We now consider a target tracking scenario:
\begin{align}
\label{simu-saut-hmc-1-tracking}
f_{k|k-1}(\x_k|\x_{k-1},r_k)&=\mathcal{N}(\x_k;\F_k(r_k)\x_{k-1};\Q_k(r_k)) \text{,} \\
\label{simu-saut-hmc-2-tracking}
g_{k}(\y_k|\x_{k},r_k)&=\mathcal{N}(\x_k;\H_k \x_{k};\R_k) \text{,}
\end{align}
where 
\begin{align*}
{\bf F}_k(r)&=\begin{bmatrix}
1 & \frac{ \sin(\omega_r T )}{\omega_r} & 0 & - \frac{1-\cos(\omega_r T)}{\omega_r} \\
0 & \cos(\omega_r T) & 0 & - \sin(\omega_r T) \\
0 & \frac{1-\cos(\omega_r T)}{\omega_r} & 1 & \frac{ \sin(\omega_r T)}{\omega_r} \\
0 &  \sin(\omega_r T) & 0 & \cos(\omega_r T)
\end{bmatrix} \text{, } \\
{\bf Q}_k(r)&= \sigma_v^2(r) \begin{bmatrix}
\frac{T^3}{3}&\frac{T^2}{2}&0&0 \\
\frac{T^2}{2}&T&0&0 \\
0&0&\frac{T^3}{3}&\frac{T^2}{2} \\
0&0&\frac{T^2}{2}&T 
\end{bmatrix} \text{, }
\end{align*}
$\H_k= {\bf I}_4$ and
${\bf R}_k= {\bf I}_4$.
We set $T=2$,
$r_k \in \{1,2,3\}$ represents
the behavior of the target: straight, left turn and right turn.
So we set $w_r=[0,6 \pi /180, -6\pi/180]$ and
$\sigma_v(r)=[7, 10, 10]$ and the 
transition probabilities are defined by
$p^1(r_{k}|r_{k-1})=0.8$ if $r_{k}=r_{k-1}$ and 
$p^1(r_{k}|r_{k-1})=0.1$ if $r_{k}\neq r_{k-1}$.

\begin{figure} [htbp!]
\center
\begin{subfigure}[]
{
\centering
\includegraphics[height=0.23 \textheight]{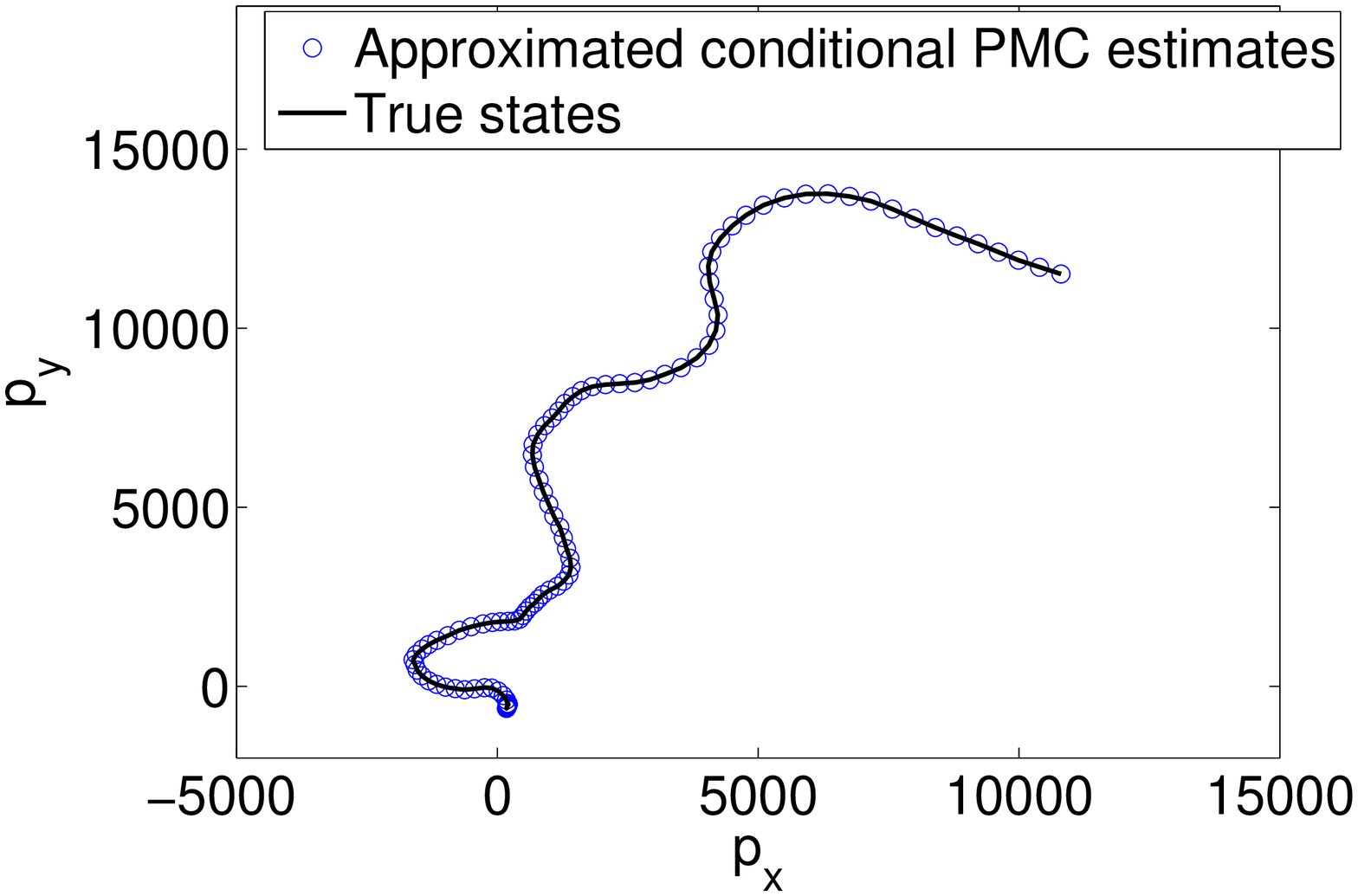} 
\label{tracking-scenario-hmc}
}
\end{subfigure} 
\begin{subfigure}[] 
{
\centering
	\includegraphics[height=0.23 \textheight]{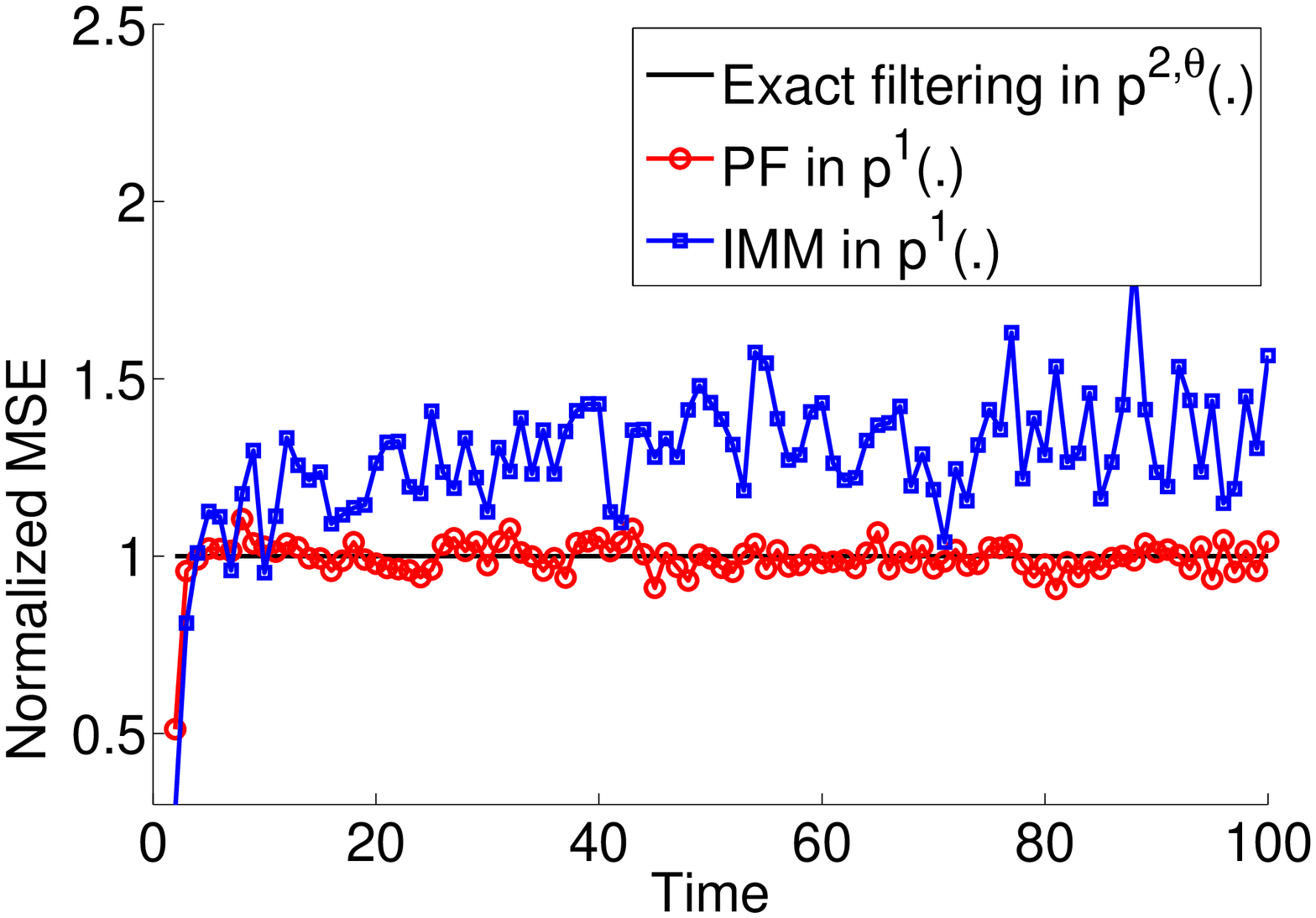} 
		\label{modele-tracking-saut-MSE}
}
\end{subfigure}
\caption{(a) - Example of a manoeuvring tracking scenario; data are generated according to model \eqref{simu-saut-hmc-1-tracking}-\eqref{simu-saut-hmc-2-tracking}.
(b) - Normalized MSE of our algorithm (black line), PF (red circles) and IMM (blue squares) estimates.}
\end{figure}

\paragraph{JMSS case} we first generate the data according
to a linear and Gaussian JMSS which satisfies
\eqref{simu-saut-hmc-1-tracking}-\eqref{simu-saut-hmc-2-tracking}.
A typical run of this manoeuvring
scenario is displayed in Fig. \ref{tracking-scenario-hmc}.
The parameters of our conditional linear and Gaussian PMC model
used to apply the exact filtering technique relies 
on the class of models of Proposition \ref{prop-2-saut}
where $\H_k^2(\r_{k-1:k})$ satisfies 
\eqref{condition-H-2} (so $\H_k^2(\r_{k-1:k})=
\F_k(r_k)$) and $\F_k^2(\r_{k-1:k})$ satisfies
\eqref{condition-F-2}.
Normalized MSE are displayed in
Fig. \ref{modele-tracking-saut-MSE}.
The solution that we have proposed 
outperforms the IMM estimate and 
presents similar performances with the PF;
however, the execution time of our algorithm
is still fifteen times faster than that of the PF.  
%%when $k \leq 50$ but the PF degrades when time increases.
%%This is because the PF approximates Gaussian Mixture 
%%$p(\x_k|\y_{0:k})$ (which grows exponentially with time)
%%by a stochastic one with $N=100$ particles, which is no
%%longer sufficient when time increases. By contrast, our approximation
%%focuses directly on the computation of $\Phi_k$ and not of that 
%%of $p(\x_k|\y_{0:k})$ which avoids the problem due 
%%to exponentially growth of this filtering pdf.

We have also averaged the MSE (w.r.t. the KF) over time and we get $0.0058$
for our solution, $0.0059$ for the PF and $0.0074$
for the IMM.

\begin{figure} [htbp!]
\centering
\begin{subfigure}[]
{
\centering
\includegraphics[height=0.23 \textheight]{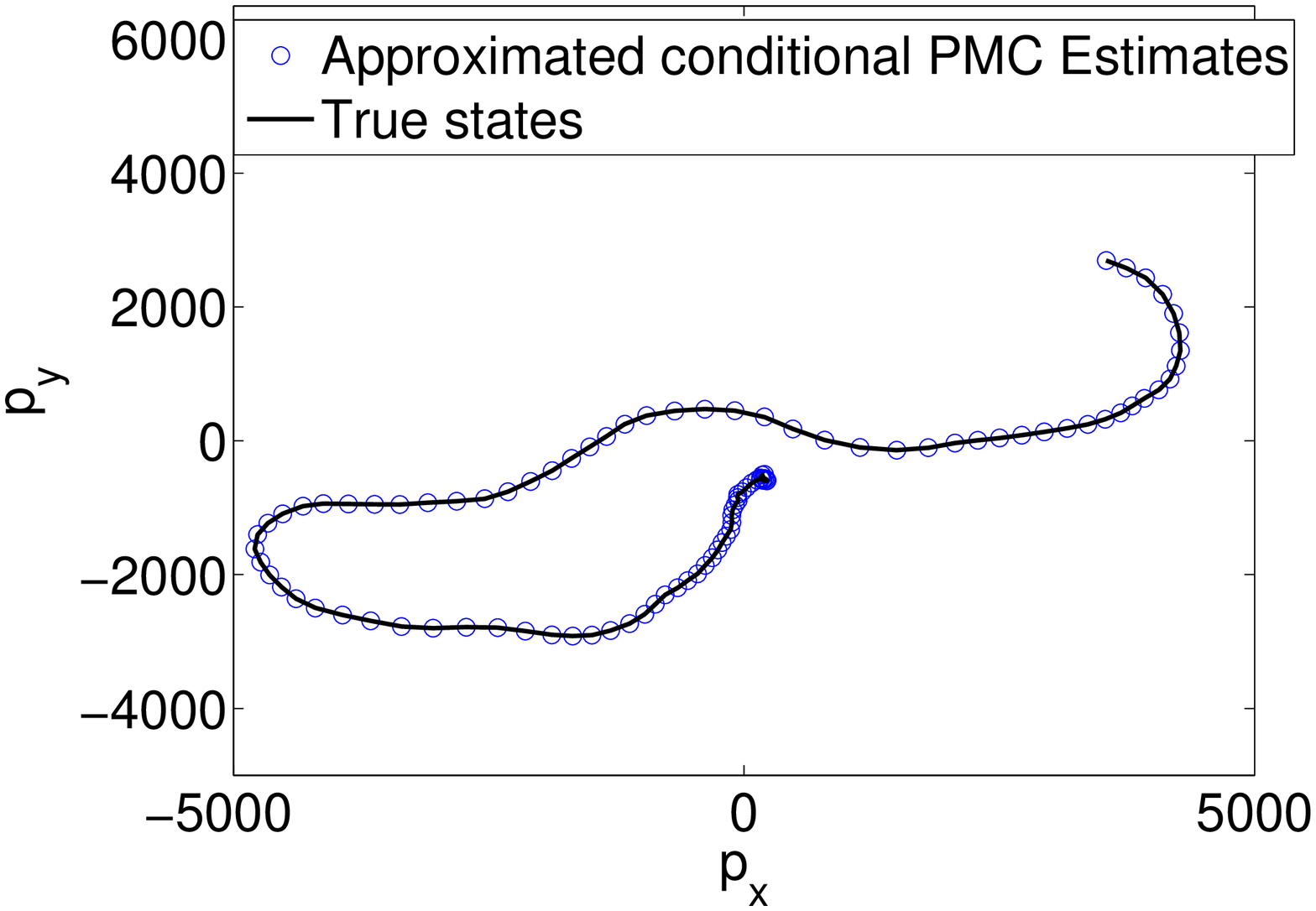} 
}
\label{tracking-scenario-pmc}
\end{subfigure} 
\begin{subfigure} []
{
\centering
\includegraphics[height=0.23 \textheight]{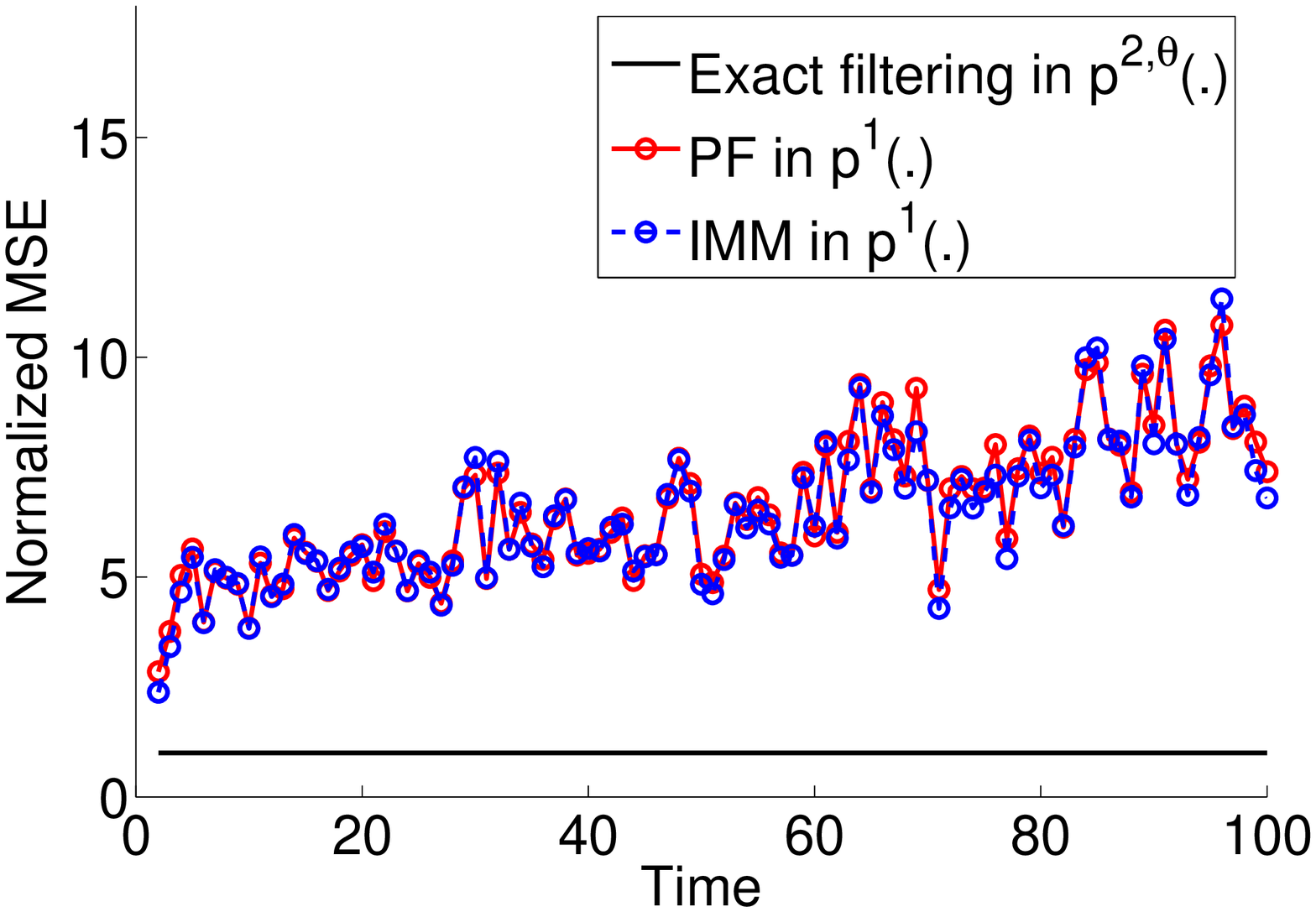} 
\label{modele-tracking-saut-pmc-MSE}
}
\end{subfigure}
\caption{ (a) - Example of a manoeuvring tracking scenario where data are now generated from a conditional linear and Gaussian PMC model with 
$\F_k^{2,{\rm true}}(\r_{k-1:k})=0.7 \F_k(r_k)$ and
$\H_k^{2,{\rm true}}(\r_{k-1:k})=0.9\H_k(r_k)$.
Physical properties of scenario of Fig. \ref{tracking-scenario-hmc} are kept.
(b) - Normalized MSE of our algorithm (black line), PF (red squares) and IMM (blue stars) estimates. Classical solutions are no longer adapted for such models while our approximation remains valid. This is because our algorithm 
offers the possibility to adjust parameter $\F_k^{2}(\r_{k-1:k})$.  }
\end{figure}

%\begin{remark}
%{\rm
\paragraph{General case}
in all these simulations, we have considered unfavorable
cases in the sense that we have generated data from linear and 
Gaussian JMSS. However, data may follow a more general statistical 
model with the same physical properties, such as model of the class
described by Proposition \ref{prop-2-saut}. However, the classical PF and IMM 
rely on the JMSS structure while our solution
is valid for a large class of models since $\F_k^2(\r_{k-1:k})$
can be arbitrary.
Let us now generate data according to a conditional PMC model
of the class described by Proposition \ref{prop-2-saut}
with $\F_k^{2,{\rm true}}(\r_{k-1:k})=0.7 \F_k(r_k)$ and
$\H_k^{2,{\rm true}}(\r_{k-1:k})=0.9\H_k(r_k)$.
We compute estimates using the same PF and IMM algorithm
that above and we compute our solution with
$\F_k^{2}(\r_{k-1:k})=0.8\F_k(r_k)$ and
$\H_k^2(\r_{k-1:k})$ satisfies
\eqref{condition-H-2}.
%so $\H_k^2(\r_{k-1:k})=\F_k(r_k)$.
Remark that setting $\F_k^{2}(\r_{k-1:k})=\F_k^{2,{\rm true}}(\r_{k-1:k})$
may not be optimal because $\H_k^2(\r_{k-1:k}) \neq \H_k^{2,{\rm true}}(\r_{k-1:k})$
and it was actually experimented that
this choice for $\F_k^{2}(\r_{k-1:k})$ gives better results.
The benchmark solution is no longer the KF since data no longer follow a JMSS
model; our reference solution is now the KF for PMC models \cite{KalmanPairwise}, 
which uses true jumps.
In Fig. \ref{tracking-scenario-pmc} we display a realization of the scenario.
As we see, 
the target keeps the physical properties of the scenario
(straight, left turn and right turn) although its trajectory
is not generated from a classical linear and Gaussian JMSS model.
However, in Fig. \ref{modele-tracking-saut-pmc-MSE} we display the normalized MSE and we see that
classical solutions are not adapted at all when we consider more 
statistical complex scenarios.

%}
%\end{remark}

%${\bf H}_k=\begin{bmatrix}
%1 & 0 & 0 & 0 \\
%0 & 0 & 1 & 0 
%\end{bmatrix}$, 
%${\bf R}_k= \sigma_{x,y} ^2\begin{pmatrix}
%\sigma_{x,y} ^2 & 0 \\
%0 & \sigma_y ^2 
%\end{pmatrix} \text{ and}$

%%%%HMC and JMSS are actually particular case of more general 
%%%%models: Pairwise Markov Chains (PMC) and Triplet Markov Chains (TMC)
%%%%models. Linear and Gaussian PMC models
%%%%generalize linear and Gaussian HMC model and exact computation 
%%%%of \eqref{Theta_n-def-generale} still holds in such models. 
%%%%So an interesting question is to study the practical interest 
%%%%of this model.
%%%%Next, TMC generalize PMC as JMSS generalize HMC: given 
%%%%$\r_{0:n}$ the TMC is a PMC. Contrary to the particular
%%%%TMC which is the JMSS, some TMC enable to compute 
%%%%\eqref{Theta_n-def-generale} without any approximation [XXX][XXX].
%%%%
%%%%The contribution of this paper is twofold:
%%%%\begin{itemize}
%%%%\item First, we focus on the interest of linear and Gaussian PMC models, and we show
%%%%that given a HMC \eqref{loi-hmc}, it is possible to derive a class of PMC 
%%%%which share the same physical properties defined by the HMC;
%%%%\item Next, we show that the previous study enables to derive 
%%%%a class of TMC model which share the same physical properties 
%%%%of a given JMSS \eqref{loi-jmss} and that it is possible in 
%%%%some case to choose a TMC in which exact filtering is 
%%%%possible at a linear computational cost.
%%%%So in this case, we have an interesting alternative to
%%%%classical solutions for the filtering problem in JMSS. 
%%%%\end{itemize}  

\section{Conclusion}
In this paper, we proposed a new filtering technique
for dynamical models with jumps.
%%which involve physical
%%properties of interest.
Starting from a given set of physical properties
we derived a class of conditional linear and Gaussian PMC models
which share those physical properties, and
in which $\Phi_k$ can be computed exactly in a computational cost
linear in the number of observations. Moreover this technique
can be used as an approximation of the MMSE estimate
in the JMSS model. 
We finally validated our approximation technique on 
simulations. Our technique provides results 
which are comparable to those given by the classical
solutions, but at a lower computational cost,
when the data is produced by a JMSS model; and 
which are better adapted in other cases.

\appendices

\section{Conditioning in random Gaussian vectors}
\label{appendice-lemma}
We recall in this section two classical results 
on Gaussian pdf which are used in our derivations \cite{rao}.
\begin{lemma}
\label{lemme-1}
Let $\zeta \in \RR^p$, ${\bf \eta} \in \RR^q$, 
$\Q$ (resp. $\P$) be a $p \times p$ (resp. $q \times q$) 
positive definite matrix  
(other vectors and matrices are of appropriate dimensions),
then
%$\eta \in \RR^q$, $F$ a $p \times q$ matrix,
%$Q$  
\begin{align}
\int \!\! \mathcal{N}(\zeta; \F {\bf \eta} \! + \! {\bf d}; \! \Q)\mathcal{N}({\bf \eta};\m;\P) {\rm d} \eta \!= \! \mathcal{N}(\zeta;\F\m \!+ \!{\bf d}; \! \Q\!+\!\F\P\F^T) \text{,}
\end{align}
\end{lemma}

\begin{lemma}
\label{lemme-2}
Let $\zeta \in \RR^p$, ${\bf \eta} \in \RR^q$, 
$\P^{\zeta}$ (resp. $\P^{\eta}$) be a $p \times p$ (resp. $q \times q$) 
positive definite matrix 
and $\P^{\zeta,\eta}$ a  $p \times q$ matrix. 
Let us assume that pdf of $(\zeta,\eta)$ is a Gaussian,
\begin{align}
p(\zeta,\eta)=\mathcal{N}(\zeta,\eta;\begin{bmatrix} \m^{\zeta} \\ \m^{\eta} \end{bmatrix}; 
\begin{bmatrix} \P^{\zeta} & {\P^{\zeta,\eta}} \\ {\P^{\zeta,\eta}}^T & \P^{\eta}  \end{bmatrix}) \text{.} 
\end{align}
Then
\begin{align}
p(\zeta,\eta)&=\mathcal{N}(\eta;\m^{\eta}; \P^{\eta})\mathcal{N}(\zeta;\tilde{\m}^{\zeta}(\eta); \tilde{\P}^{\zeta}), \\
\tilde{\m}^{\zeta}(\eta)&= \m^{\zeta}+ {\P^{\zeta,\eta}}(\P^{\eta})^{-1}(\eta-\m^{\eta} ) \text{,} \\
\tilde{\P}^{\zeta}&= \P^{\zeta}-\P^{\zeta,\eta}(\P^{\eta})^{-1}{\P^{\zeta,\eta}}^T \text{.}
 \end{align}
\end{lemma}

\section{Proof of Propositions \ref{remark-1} to \ref{remark-3}} 
\label{proof-remark-2}
We begin with \eqref{loi-sachant-x-k-1}.
Let $p^{2,\theta}(\x_k,\y_k|\x_{k-1},\y_{k-1})$ 
be the transition pdf
of a PMC model of Proposition \ref{prop-1}.
We have 
\begin{align}
\nonumber
p^{2,\theta}(\x_k,\y_k|\x_{k-1})&= 
\int 
\underbrace{p^{2,\theta}(\y_{k-1}|\x_{k-1})}_{g_{k-1}(\y_{k-1}|\x_{k-1})}
%p^{2,\theta}_{k|k-1}(\z_k|\z_{k-1})  {\rm d}\y_{k-1}
%\times \nonumber \\ &
p^{2,\theta}_{k|k-1}(\x_k,\y_k|\y_{k-1},\x_{k-1})  {\rm d}\y_{k-1}.
\end{align}
%%%%%%%%%%%%%\begin{align}
%%%%%%%%%%%%%\nonumber
%%%%%%%%%%%%%p^2(\x_k,\y_k|\x_{k-1})&= \int p^2_{k|k-1}(\x_k,\y_k|\y_{k-1},\x_{k-1}) p^2(\y_{k-1}|\x_{k-1}) {\rm d}\y_{k-1} \\
%%%%%%%%%%%%%\nonumber
%%%%%%%%%%%%%&= \int p^2_{k|k-1}(\x_k,\y_k|\y_{k-1},\x_{k-1}) g_{k-1}(\y_{k-1}|\x_{k-1}) {\rm d}\y_{k-1} .
%%%%%%%%%%%%%\end{align}
Now $g_{k-1}(\y_{k-1}|\x_{k-1})\!\!= \!\! \mathcal{N}(\y_{k-1} \! ;\H_{k-1}\x_{k-1} \! ;$ $\R_{k-1})$
and
$p^{2,\theta}_{k|k-1}(\x_k,\y_k|\y_{k-1},\x_{k-1})$ is a Gaussian 
given by parameters \eqref{matrice-B}-\eqref{matrice-sigma-22}.
Using Lemma \ref{lemme-1}, we get \eqref{loi-sachant-x-k-1}.
We now prove \eqref{loidey_ksachantx_0:k} by induction.
So let us assume that
\begin{equation}
\label{recurrence}
p^{2,\theta}(\y_{k-1}|\x_{0:k-1}) =p^{2,\theta}(\y_{k-1}|\x_{k-1})= g_{k-1}(\y_{k-1}|\x_{k-1})
\end{equation}
(\eqref{recurrence} is true at time $k=1$).
Since $(\x_{0:k},\y_{0:k})$ is a PMC, we get successively
\begin{eqnarray}
\nonumber
p^{2,\theta}(\x_k,\y_k|\x_{0:k-1})
 \! \!\!\!\!\! & \stackrel{\rm PMC}{=} & \!\!\!\!\!\!
\int p^{2,\theta}_{k|k-1}(\z_k|\z_{k-1},\y_{k-1})
%%\times \nonumber \\ &&
p^2(\y_{k-1}|\x_{0:k-1}) {\rm d} \y_{k-1} 
\\
\label{dzedz}
& \stackrel{\eqref{recurrence}}{=} &
p^{2,\theta}(\z_k|\x_{k-1})
\nonumber \\ &
\stackrel{\eqref{loi-sachant-x-k-1}}{=} &
f_{k|k-1}(\x_k|\x_{k-1}) g_k(\y_k|\x_k).
\end{eqnarray}
From \eqref{dzedz} we get
\begin{eqnarray}
\label{demo-x-markov}
p^{2,\theta}(\x_k|\x_{0:k-1})
& = &
f_{k|k-1}(\x_k|\x_{k-1}),
\end{eqnarray}
and consequently
$p^{2,\theta}(\y_k|\x_{0:k}) = g_k(\y_k|\x_k)$, 
which is nothing but \eqref{recurrence} at time $k$,
which proves \eqref{loidey_ksachantx_0:k}.
Now since \eqref{recurrence} is true 
\eqref{demo-x-markov} holds too,
whence \eqref{p2xestmarkov}.
It remains to prove \eqref{pysachantx-caspmc}.
Let ${\cal N}$ stand for numerator. 
Since $\{ (\x_k,\y_k) \}_{n\geq 0}$ is a MC,
$p^2(\y_i|\y_{0:i-1},\x_{0:k}) =$
$\frac{p^2(\y_{0:i},\x_{0:k})}{\int p^2(\y_{0:i},\x_{0:k}){\rm d}\y_i} =$
$\frac{p^2(\x_{i:k},\y_{i}|\x_{i-1},\y_{i-1}) p^2(\x_{0:i-1},\y_{0:i-1})}{\int {\cal N}{\rm d}\y_i} =$
$p^2(\y_i|\y_{i-1},\x_{i-1:k})$,
whence \eqref{pysachantx-caspmc}.

\section{Proof of Proposition \ref{exact-filtering}}
\label{proof-exact-filtering}
The results is based on the filtering 
technique in \cite{wp-cras-exact}.
We consider a conditional linear and Gaussian PMC
model which satisfies  
\eqref{condition-filtering}-\eqref{loi-conditionnelle-saut-1}
and the goal is to compute 
$p^2(r_{k}|\y_{0:k})$ from $p^2(r_{k-1}|\y_{0:k-1})$ and
${\rm E}({\bf x}_{k}|\y_{0:k},r_{k})$ from  ${\rm E}({\bf x}_{k-1}|\y_{0:k-1},r_{k-1})$.
In this particular conditional PMC model,
$(\y_{0:k-1},\r_{0:k-1})$ is a MC \cite{wp-cras-exact}, so
\begin{align}
\label{r-y-MC}
p^2(\y_{k},r_{k} |\y_{0:k-1},r_{k-1}) &= p^2(\y_{k},r_{k}|\y_{k-1},r_{k-1})\text{,} \\
&= p^2(r_{k}|r_{k-1}) p^2(\y_{k}|\y_{k-1},\r_{k-1:k})
\end{align}
Consequently,
\begin{align}
p^2(r_{k}|\y_{0:k}) &\propto 
%\frac{ \sum_{r_k} p^2(r_{k+1}|r_k) p^2(\y_{k+1}|\y_k,\r_{k:k+1})}
%{ \sum_{r_{k+1},r_k} p^2(r_{k+1}|r_k) p^2(\y_{k+1}|\y_k,\r_{k:k+1})} \text{.}
\sum_{r_{k-1}} p^2(r_{k-1}|\y_{0:k-1}) p^2(\y_{k},r_{k}|\y_{k-1},r_{k-1})
\text{.}
\end{align}
Next, 
\begin{align}
\label{esperance-conditionnelle-proof}
&{\rm E}({\bf x}_{k}|\y_{0:k},r_{k})= \sum_{r_{k-1}} p^2(r_{k-1}|\y_{0:k},r_{k}) 
\times \nonumber \\ &
\int \underbrace{\left[\int \x_{k} p^2(\x_{k}|\x_{k-1},\y_{0:k},\r_{k-1:k}) {\rm d}\x_{k} \right]}_{{\rm E}(\x_{k}|\x_{k-1},\y_{0:k},\r_{k-1:k})} p^2(\x_{k-1}|\y_{0:k},\r_{k-1:k}){\rm d}\x_{k-1} \text{,}
\end{align}
Let us now compute the quantities involved in \eqref{esperance-conditionnelle-proof} 
From \eqref{r-y-MC}, we have
\begin{align}
\label{quantite-1}
p^2(r_{k-1}|\y_{0:k},r_{k}) & \propto p(r_{k-1}|\y_{0:k-1})p(\y_{k},r_{k}|\y_{k-1},r_{k-1}) \text{.} 
\end{align}
Because $(\x_{0:k},\y_{0:k},\r_{0:k})$ is a MC,
\begin{align}
\label{quantite-2}
p^2(\x_{k}|\x_{k-1},\y_{0:k},\r_{k-1:k})&=p^2(\x_{k}|\x_{k-1},\y_{k-1},\y_{k},\r_{k-1:k}) \text{,}
\end{align}
so from \eqref{loi-conditionnelle-saut-1} we deduce 
\begin{align}
\label{quantite-2-prime}
{\rm E}(\x_{k}|\x_{k-1},\y_{0:k},\r_{k-1:k})&=
{\bf C}_{k}(\r_{k-1:k}) \x_{k-1}
%\nonumber \\ &
+{\bf D}_{k}(\r_{k-1:k},\y_{k-1:k})
\end{align}
Next, in this particular conditional PMC model,
\begin{align}
\label{quantite-3}
p^2(\x_{k-1}|\y_{0:k},\r_{k-1:k})&= p^2(\x_{k-1}|\y_{0:k-1},r_{k-1}) \text{.}
\end{align}
Finally, plugging \eqref{quantite-1},\eqref{quantite-2-prime}
and \eqref{quantite-3} in \eqref{esperance-conditionnelle-proof},
we get \eqref{exact-filtering-esperance}.
The proof for the computation of
${\rm E}({\bf x}_{k}\x_{k}^T|\y_{0:k},r_{k})$
is similar.

\section{Proof of Proposition \ref{proposition-KL}}
\label{proof-prop-4}
Let us consider the class of conditionl linear and Gaussian PMC models
of Proposition \ref{prop-2-saut} which satisfy \eqref{condition-H-2}.
We compute the KLD ${\rm D}_{\rm KL}(p^{2,\theta}(\z_{0:k},\r_{0:k}),p^1(\z_{0:k},\r_{0:k}))$
which can be rewritten 
\begin{align}
{\rm D}_{\rm KL}(p^{2,\theta}&(\z_{0:k},\r_{0:k}),p^1(\z_{0:k},\r_{0:k})) = 
\sum_{\r_{0:k}} p^1(\r_{0:k}) 
%\times \nonumber \\ &
{\rm D}_{\rm KL}(p^{2,\theta}(\z_{0:k}|\r_{0:k}),p^1(\z_{0:k}|\r_{0:k})) 
\end{align}
because $p^1(\r_{0:k})=p^{2,\theta}(\r_{0:k})$ (see Proposition \ref{proposition-invariance-saut}).
$p^1(\r_{0:k})$ does not depend on $\{\F_k^2(\r_{k-1:k})\}_{k \geq 1} $, so 
we focus on ${\rm D}_{\rm KL}(p^{2,\theta}(\z_{0:k}|\r_{0:k}),p^1(\z_{0:k}|\r_{0:k}))$. 
Using Markovian properties, we have
\begin{align}
&{\rm D}_{\rm KL}(p^{2,\theta}(\z_{0:k}|\r_{0:k}),p^1(\z_{0:k}|\r_{0:k}))
\!\! = \!\! 
\sum_{j=1}^k \int p^{2,\theta} (\z_{j-1}|\r_{0:j-1})
\times \nonumber \\ &
{\rm D}_{\rm KL}(p^{2,\theta}_{j|j-1}(\z_j|\z_{j-1},\r_{j-1:j}),p^1_{j|j-1}(\z_j|\z_{j-1},\r_{j-1:j})) {\rm d}\z_{j-1} \text{,}
\end{align}
where, according to Propositions \ref{prop-2-saut} and \ref{proposition-invariance-saut},
$p^{2,\theta} (\z_{j-1}|\r_{0:j-1})=$ $p^{1} (\z_{j-1}|\r_{0:j-1})$ 
and so does not depend on $\F_j^2(\r_{j-1:j})$.
So we just minimize ${\rm DKL}(p^{2,\theta}_{j|j-1}(\z_j|\z_{j-1},\r_{j-1:j}),p^1_{j|j-1}(\z_j|\z_{j-1},\r_{j-1:j})).$
We have
%%According to the Bayes'rule,
%%$p^1(\z_j|\z_{j-1},\r_{j-1:j})=p^1(\x_j|\x_{j-1},\y_{j},r_j)p^1(\y_j|\x_{j-1},r_j)$
%%and $p^{2,\theta}_{j|j-1}(\z_j|\z_{j-1},\r_{j-1:j})=$ $p^{2,\theta}(\x_j|\z_{j-1},\y_{j},\r_{j-1:j})p^{2,\theta}(\y_j|\y_{j-1},\r_{j-1:j})$
%%(remember that $p^{2,\theta}(.)$ satisfies \eqref{condition-H-2})
%%where
\begin{align}
\label{moyenne-p-2-H-2}
p^{2,\theta}(\y_j|\y_{j-1},\r_{j-1:j}) & =  \mathcal{N}(\y_j; \H_j^2(\r_{j-1:j}) \y_{j-1};  \R_{j}(r_j)- 
% \nonumber \\ &
\H^2_{j}(\r_{j-1:j})\R_{j-1}(r_{j-1}){\H^2_j(\r_{j-1:j})}^T 
 \nonumber \\ &
+  \H_{j}(r_j)\Q_{j}(r_j)\H_{j}(r_j)^T) \text{,} \\
%\end{align}
%and
%\begin{align}
p^{2,\theta}(\x_j|\z_{j-1},\y_j,\r_{j-1:j})&= \mathcal{N}(\x_j; \m_j^{\x_j};\P_j^{\x_j}) \text{,} \\
%\end{align}
%with 
%\begin{align}
\label{moyenne-p-2-F-2}
\m_j^{\x_j} \!\!& =\!\! (\F_j(r_j)\!-\!\F_j^2(\r_{j-1:j})\! \H_{j-1}(r_{j-1}))\x_{j-1} \!\! +\!\! \F_j^2(\r_{j-1:j})\y_{j-1} + 
\nonumber \\ &
+ ({\bf \Sigma}_j^{21}(\r_{j-1:j}))^T \! ({\bf \Sigma}_j^{22}(\r_{j-1:j}))^{-1}\! (\y_j\!\!-\!\! \H_j^2(\r_{j-1:j}) \y_{j-1} \!) \text{,} \\
\P_j^{\x_j}&= {\bf \Sigma}_j^{11}(\r_{j-1:j}) - ({\bf \Sigma}_j^{21}(\r_{j-1:j}))^T 
%\times \nonumber \\ &
({\bf \Sigma}_j^{22}(\r_{j-1:j}))^{-1}{{\bf \Sigma}_j^{21}(\r_{j-1:j})} \text{,}
\end{align}
where ${\bf \Sigma}_j^{11}(\r_{j-1:j})$, 
${\bf \Sigma}_j^{21}(\r_{j-1:j})$ and ${\bf \Sigma}_j^{22}(\r_{j-1:j})$
are defined in \eqref{sigma-11-saut}-\eqref{sigma-22-saut}.
Next, the KLD between $p^{2,\theta}_{j|j-1}(\z_j|\z_{j-1},$ $\r_{j-1:j})$ 
and $p^{1}_{j|j-1}(\z_j|\z_{j-1},\r_{j-1:j})$
writes as
\begin{align}
{\rm D}_{\rm KL}(p^{2,\theta}_{j|j-1},p^1_{j|j-1}) &= \int p^{2,\theta}_{j|j-1}(\z_j|\z_{j-1},\r_{j-1:j})
%\times \nonumber \\ &
 \log \left( \frac{p^{2,\theta}_{j|j-1}(\z_j|\z_{j-1},\r_{j-1:j})}{p^1_{j|j-1}(\z_j|\z_{j-1}),\r_{j-1:j}}\right) {\rm d} \z_j \text{,} \\
&= {\rm D}_{\rm KL}(p^{2,\theta}(\y_j|\y_{j-1},\r_{j-1:j}),p^1(\y_j|\x_{j-1},r_j)) 
+ 
%\nonumber \\ &
\int p^{2,\theta}(\y_j |\y_{j-1},\r_{j-1:j})
\times \nonumber \\ &
 {\rm D}_{\rm KL}(p^{2,\theta}(\x_j|\z_{j-1},\y_j,\r_{j-1:j}),p^1(\x_j|\x_{j-1},\y_j,\r_{j-1:j})) {\rm d}\y_j
\end{align}
and is minimum when 
$p^{2,\theta}(\x_j|\z_{j-1},\y_j,\r_{j-1:j})=p^1(\x_j|\x_{j-1},\y_j,r_{j})$
(from \eqref{moyenne-p-2-H-2}, $p^{2,\theta}(\y_j |\y_{j-1},\r_{j-1:j})$ does not
depend on $\F_j^2(\r_{j-1:j})$).
From Proposition \ref{proposition-invariance-saut}, we know that
$$p^{2,\theta}(\x_j|\x_{j-1},\y_j,\r_{j-1:j})=p^1(\x_j|\x_{j-1},\y_j,\r_{j-1:j})$$
so ${\rm D}_{\rm KL}(p^{2,\theta}_{j|j-1}(\z_j|\z_{j-1},\r_{j-1:j}),p^1_{j|j-1}(\z_j|\z_{j-1},\r_{j-1:j}))$ is minimum
when $p^2(\x_j|\z_{j-1},\y_j,\r_{j-1:j})$ does not 
depend on $\y_{j-1}$. From \eqref{moyenne-p-2-F-2},
classical calculus lead to
\begin{align}
&\F_j^{2}(\r_{j-1:j})= \Q_j(r_j) \H_j(r_j)^T 
%\times \nonumber \\ &
\left [\R_j(r_j) + \H_j(r_j) \Q_j(r_j) \H_j(r_j)^T \right]^{-1} \H_j^2(\r_{j-1:j}) \text{.}
\end{align}
%\begin{align}
%p^2(\x_k|\x_{k-1},\y_k)&=  \frac{ p(\x_k|\x_{k-1})p(\y_k|\x_k,\x_{k-1})} { \int p(\x_k|\x_{k-1})p(\y_k|\x_k,\x_{k-1}) {\rm d} \x_k} \\
%&= \frac{ f_{k|k-1}(\x_k|\x_{k-1})g_k(\y_k|\x_k)} { \int f_{k|k-1}(\x_k|\x_{k-1}) g_k(\y_k|\x_k) {\rm d} \x_k} \\
%&= p^1(\x_k|\x_{k-1},\y_k)
%\end{align}

% or


%\appendices
%% you can choose not to have a title for an appendix
%% if you want by leaving the argument blank
%\section{}

%%\section*{Acknowledgment}
%%
%%
%%The authors would like to thank...

\ifCLASSOPTIONcaptionsoff
  \newpage
\fi

\bibliographystyle{IEEEtran}%%\begin{align}
%%{\rm D}_{\rm KL}(p^2(\z_{0:k},\r_{0:k}),p^1(\z_{0:k},\r_{0:k}))&= \sum_{\r_{0:k}} p^1(\r_{0:k}) {\rm D}_{\rm KL}(p^2(\z_{0:k}|\r_{0:k}),p^1(\z_{0:k}|\r_{0:k}))
%%\end{align}

\bibliography{yohan-p-these}

\end{document}